\documentclass[showpacs,preprintnumbers,amsmath,amssymb,prb]{revtex4}

\usepackage{graphicx}
\usepackage{dcolumn}
\usepackage{bm}

\begin{document}

\title{Frustrated impurity spins in ordered two-dimensional quantum antiferromagnets}

\author{A. V. Syromyatnikov}
 \email{syromyat@thd.pnpi.spb.ru}
\author{S. V. Maleyev}
\affiliation{Petersburg Nuclear Physics Institute, Gatchina, St.\ Petersburg 188300, Russia}

\date{\today}

\begin{abstract}

Dynamical properties of an impurity spin coupled {\it symmetrically} to sublattices of ordered 2D Heisenberg quantum antiferromagnet (i.e., frustrated impurity spin) are discussed at $T\ge0$ (existence of a small interaction stabilizing the long range order at $T\ne0$ is implied). We continue our study on this subject started in Phys.\ Rev.\ B {\bf 72}, 174419 (2005), where spin-$\frac12$ defect is discussed and the host spins fluctuations are considered within the spin-wave approximation (SWA). In the present paper we i) go beyond SWA and ii) study impurities with spins $S \ge 1/2$. It is demonstrated that in contrast to defects coupled to sublattices asymmetrically longitudinal host spins fluctuations play important role in the frustrated impurity dynamics. We show that the effect of the host system on the defect is completely described by the spectral function as it was within SWA. The spectral function, that is proportional to $\omega^2$ within SWA, acquires new terms proportional to $\omega^2$ and $\omega T^2$ originating from longitudinal host spins fluctuations. It is observed that the spin-$\frac12$ impurity susceptibility has the same structure as that obtained within SWA: the Lorenz peak and the non-resonant term. The difference is that the width of the peak becomes larger being proportional to $f^2(T/J)^3$ rather than $f^4(T/J)^3$, where $f$ is the dimensionless coupling parameter. We show that transverse static susceptibility acquires a new negative logarithmic contribution.
In accordance with previous works we find that host spins fluctuations lead to an effective one-ion anisotropy on the impurity site. Then defects with $S>1/2$ appears to be split. We observe strong reduction of the value of the splitting due to longitudinal host spins fluctuations. We demonstrate that the dynamical impurity susceptibility contains $2S$ Lorenz peaks corresponding to transitions between the levels, and the non-resonant term. The influence of finite concentration of the defects $n$ on the low-temperature properties of antiferromagnet is also investigated. Strong spin-wave damping proportional to $nf^4|\omega|$ is obtained originating from the non-resonant terms of the susceptibilities. 

\end{abstract}


\maketitle

\section{Introduction}

\begin{figure}
\centering
\includegraphics[scale=0.5]{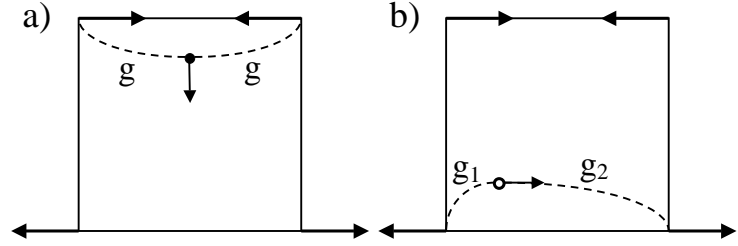}
\caption{Unit cell of 2D AF with impurity spin coupled (a) symmetrically and (b) asymmetrically to AF sublattices. Strengths of coupling with corresponding host spins $g$ and $g_1\ne g_2$ are depicted. The local Neel order is also shown. Only symmetrically coupled impurities are discussed in the present paper.
\label{pic}} 
\end{figure}

The problem of impurity spins in 2D Heisenberg quantum antiferromagnet (AF) has been attracted much attention in last two decades. \cite{wan,hog1,hog2,sush1,sush2,sachdev,vojta,nagaosa,igar,murayama,clarke,oitmaa,kot2,i} The majority of works are devoted to defects coupled {\it asymmetrically} to sublattices of AF. Among such impurities are added spin coupled to one host spin, substitutional spin and vacancy that is the particular case of the added spin with the coupling strength $g\to\infty$. Such works are stimulated by a variety of experiments studying cuprates with magnetic and nonmagnetic impurities. Defects coupled {\it symmetrically} to sublattices of AF (see Fig.~\ref{pic}) are much less studied because such objects are much rare in occurrence. At the same time they have different properties \cite{clarke,i} and are also of interest. For instance, attempts were made to model holes in CuO-planes of some high-$T_c$ compounds being in antiferromagnetic phase by spin-$\frac12$ impurity coupled symmetrically to two neighboring host spins. \cite{clarke,oitmaa,kot2} 

In our recent paper \cite{i} (hereafter referred to as I) we studied dynamical properties of impurity coupled symmetrically to two neighboring spins in 2D AFs at $T\ge0$. A technique was proposed based on Abrikosov's pseudofermion technique \cite{abrikos} that allows to discuss dynamics of an impurity with arbitrary spin value $S$. Meanwhile we focus in I on the particular case of spin-$\frac12$ defect. Our aim was to find the impurity dynamical susceptibility $\chi(\omega)$. We assumed that the impurity dynamics is governed by interaction with spin waves. Then, two kinds of the host systems are considered in I: i) ordered 2D AF in which the long range order at $T\ne0$ is stabilized by a small interaction (for definiteness interplane interaction) $\eta\ll J$, where $J$ is the coupling constant between the host spins; ii) isotropic 2D AF. We bared in mind that in isotropic 2D AF only spin waves are well defined with energies much larger than $Jsa/\xi$, where $a$ is the lattice constant, $s$ is the value of host spin, and $\xi\propto \exp({\rm const}/T)$ is the correlation length. It was obtained within the spin-wave approximation (SWA) that, similar to the spin-boson model, the effect of the host system on the defect is completely described by the spectral function which was assumed to be proportional to $\omega^2$ in calculations of $\chi(\omega)$. We demonstrated that the spectral function is proportional to $\omega^2$ in 2D AF and in the ordered quasi-2D AF at $\omega\gg Jsa/\xi$ and $\omega\gg\eta$, respectively. Notice that only transverse host spins fluctuations contribute to the spectral function within SWA. Then, only processes of absorption/emission of one spin wave by the impurity is taken into account within SWA.

The calculations of $\chi(\omega)$ were performed in I within the fourth order of the dimensionless coupling parameter $f\propto g/J$. We shown that the transverse impurity susceptibility $\chi_\perp(\omega)$ has a Lorenz peak with the width $\varGamma\propto f^4J(T/J)^3$ that disappears at $T=0$, and a non-resonant term. The imaginary part of the non-resonant term is a constant independent of $T$ at $|\omega|\gg\varGamma$ and the real part has a logarithmic divergence of the form $f^2\ln([\omega^2 + \varGamma^2]/J^2)$. The longitudinal susceptibility $\chi_\|(\omega)$ has only the non-resonant term which differs from that of $\chi_\perp(\omega)$ by a constant. The fact that the spectral function in 2D AF is proportional to $\omega^2$ only at $\omega\gg\{\eta \mbox{ \rm or } Jsa/\xi\}$ leads to the following restriction on the range of validity of the results obtained: $\max\{\varGamma,|\omega|\}\gg\{\eta \mbox{ \rm or } Jsa/\xi\}$. Notice that in the case of isotropic 2D AF at $T\ne0$ this relation determined the range of validity of our assumption that the impurity dynamics is governed by interaction with spin waves. 

It was observed that the static susceptibility $\chi(0)$ has the free-spin-like term $1/(4T)$ and a correction proportional to $f^2\ln (J/T)$. Quite a different behavior of $\chi(0)$ was obtained in the regime of $T\ll |g|$ for asymmetrically coupled defects: \cite{sush2,sachdev} the classical-like term $S^2/(3T)$ and the logarithmic correction proportional to $\ln (J/T)$ rather than  $f^2\ln (J/T)$. 

The influence of the finite concentration of the defects $n$ on the low-temperature properties of 2D AF was also studied in I. It was shown that correction to the square of the spin-wave spectrum is proportional to $nf^2\chi_\perp(\omega)$. Then two regimes were considered,  $|\omega|\gg\omega_0\propto f^2T^2/J$ and $|\omega|\ll\omega_0$, at which, respectively, non-resonant and resonant parts of $\chi_\perp(\omega)$ prevail. In the most interesting case of $|\omega|\gg\omega_0$ we found the logarithmic correction to the spin-wave velocity and an anomalous damping of the spin-waves proportional to $nf^4|\omega|$. Such damping was obtained also in Refs.~\cite{chern,wan} where asymmetrically coupled impurities were studied in 2D AF. 

In the present paper we continue our discussion of symmetrically coupled defects in 2D AFs and i) go beyond SWA and ii) study defects with $S\ge1/2$. In particular, longitudinal host spins fluctuations come into play outside SWA. They lead to two kinds of processes: absorption/emission of two magnons by the impurity and scattering of one spin wave on the impurity. We find that in the case of isotropic 2D AF at $T\ne0$ our assumption that the impurity dynamics is governed by interaction with spin waves is wrong at any $T$ and $\omega$. This circumstance manifests itself in the fact that spin waves with wavelength of the order of $\xi$ are important in the processes of absorption/emission of two magnons and scattering of one magnon. Thus, our approach can be applied to ordered 2D AFs only. 

It is demonstrated that longitudinal host spins fluctuations lead to two important contributions to the spectral function which should be taken into account. One of them is proportional to $T^2\omega/s$ and corresponds to scattering of one spin wave on the impurity. Another contribution has the form $\omega^2v(T)/s$, where $v(T)$ is a series in $(T/s)\ln(T/[s\eta])$, and originates from two processes: emission or absorption of two spin waves by the impurity and the scattering of one magnon on the impurity.

It is demonstrated that spin-$\frac12$ impurity susceptibility $\chi_\perp(\omega)$ has the same structure as that obtained within SWA: the Lorenz peak and the non-resonant term. The difference is that the width of the peak $\varGamma$ acquires large correction of the first order of $f^2$: $\varGamma \propto f^2J(T/J)^3$. This term dominates if the following condition is fulfilled that does not depend on $s$: $|g|/J \ll 5$. Besides, some constants in the expression for $\chi_\perp(\omega)$ acquire thermal corrections. The width of Lorenz peak in $\chi_\|(\omega)$ is zero within our precision. We observe that longitudinal host spins fluctuations lead to a new logarithmic correction to $\chi_\perp(0)$ proportional to $-(f^2/[Js])\ln(T/[s\eta])$.

Corrections to the spin-wave spectrum have the same form as those obtained within SWA. The difference is that some constants acquire thermal corrections, $\varGamma$ has another form and the quantity $\omega_0$ separating two regimes becomes larger being proportional to $T^2/J$ rather than $f^2T^2/J$.

It should be pointed out that the obtained remarkable influence of longitudinal host spins fluctuation on frustrated impurity spin dynamics is quite unusual. There is no such influence in the case of asymmetrically coupled defects. We show below that longitudinal fluctuations give a negligibly small correction to the spectral function and there is no reason to expect any perceptible influence from them. Thus, we confirm that the frequently used $T$-matrix approach \cite{iz,wan,chern} basing on Dyson-Maleev transformation and dealing with only bilinear part of the Hamiltonian is appropriate for discussing the asymmetrically coupled defects.

We find also $\chi(\omega)$ for symmetrically coupled impurities with $S>1/2$. It is well-known that host spins fluctuations lead to an effective one-ion anisotropy on the impurity site in 3D AF: $-\tilde {\cal C}(T)S_z^2g^2/J$, where $\tilde {\cal C}(T)>0$. \cite{shender,ivanov} Then defects with $S>1/2$ appear to be split and magnetization of sublattices is a hard axis for them. We obtain the same one-ion anisotropy for defects in 2D AF. In particular we observe large reduction of the value of $\tilde {\cal C}(T)$ obtained within SWA by $1/s$-corrections stemming from longitudinal host spins fluctuations. We show that thermal corrections reduce $\tilde {\cal C}(T)$ further and can change the sign of $\tilde {\cal C}(T)$ at $T\alt T_N$.

Then we find that if $T$ is greater than the value of the defect splitting, transverse dynamical susceptibility contains $2S$ Lorenz peaks corresponding to transitions between impurity levels and a nonresonant contribution. Widths of the peaks are proportional to $f^2J(T/J)^3$ as in the case of $S=1/2$. When $T$ is lower than the distance between nearest impurity levels ($\sim \tilde {\cal C}(T)g^2/J$) there are only peaks corresponding to transitions between low-lying levels. Longitudinal susceptibility has only the nonresonant term. At $T\gg \tilde {\cal C}(T)g^2S^2/J$ static susceptibility has the same structure as that of $S=1/2$: term $S(S+1)/(3T)$ and a logarithmic correction. If $T\ll \tilde {\cal C}(T)g^2/J$, $\chi(0)$ of integer and half-integer spins behave differently. The imaginary part of the nonresonant term is a constant that leads to abnormal spin-wave damping proportional to $nf^4|\omega|$, as in the case of spin-$\frac12$ impurity.

It is explained in I that in the case of finite concentration of defects the results obtained within our method have a limited range of validity. This is because the interaction with defects modifies host spins Green's functions and the bare spectral function acquires large corrections at energies around those of the Lorenz peaks. The problem should be solved self-consistently to obtain correct expressions for the impurity susceptibility and the spin-wave spectrum around the resonance energies. Corresponding consideration is out of the scope of the present paper. This large renormalization of the spectral function can be interpreted as a resonance scattering of spin waves on impurities. Similar situation exists in the case of asymmetrically coupled defects (see discussion in Sec.~\ref{influence}). As a result in the vicinity of the resonance energies one might expect strong deviation from linearity of the real part of the spin-wave spectrum and an increase of the spin-wave damping.

The rest of this paper is organized as follows. The model and diagrammatic technique are discussed in Sec.~\ref{model}. Pseudofermion Green's functions and pseudofermion vertex are calculated in Sec.~\ref{gfver}. The impurity dynamical susceptibility is derived in Sec.~\ref{ds}. Influence of the defects on the spin-wave spectrum is considered in Sec.~\ref{influence}. Section~\ref{concl} contains our conclusions. There are two appendixes with details of calculations.

\section{Technique and general expressions}
\label{model}

\subsection{Technique}

\begin{figure}
\centering
\includegraphics{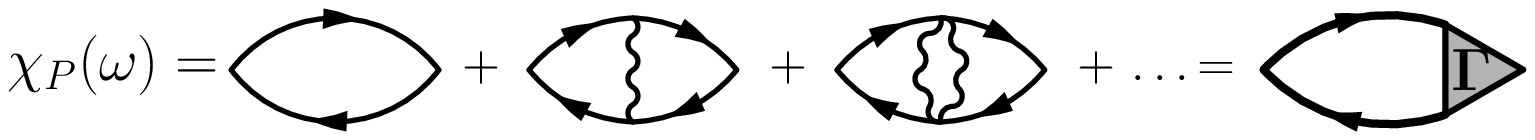}
\caption{Lower-order diagrams for the impurity dynamical susceptibility $\chi_P(\omega)$ and a graphical representation of the result of the overall series summation. Lines with arrows and wavy lines represent pseudofermion and magnons Green's functions, respectively.
\label{chifig}} 
\end{figure}

We discuss Heisenberg AF with impurity spin $\bf S$ coupled to two neighboring host spins ${\bf s}_1$ and  ${\bf s}_2$ describing by the Hamiltonian
\begin{equation}
\label{ham}
{\cal H} = J\sum_{\langle i,j\rangle}{\bf s}_i {\bf s}_j + g {\bf S} ({\bf s}_1 + {\bf s}_2).
\end{equation}
It should be stressed that one can consider other symmetrically coupled defects (e.g., those coupled to four host spins) on the equal footing. The results would differ by some constants only. It is convenient to use Abrikosov's pseudofermion technique \cite{abrikos} and to represent the impurity spin as ${\bf S} = \sum_{nm}b^\dagger_n{\bf S}_{nm}b_m$, where $n$ and $m$ are the spin projections, $b^\dagger_n$ and $b_n$ are operators of creation and annihilation of pseudofermions. We calculate below impurity susceptibility $\chi(\omega)$ using diagrammatic technique. 
First diagrams for $\chi(\omega)$ and a graphical representation of the result of all diagrams summation are shown in Fig.~\ref{chifig}. Thin lines with arrows in the picture represent the bare pseudofermion Green's functions: $G_{mm'}^{(0)} (i\omega_n) =\delta_{mm'}(i\omega_n-\lambda)^{-1}$, where $\lambda$ is the chemical potential of pseudofermions that should be tended to infinity in the resultant expressions. Wavy lines in the picture denote magnons Green's functions. As usual, diagrams with only one pseudofermion loop should be taken into account because each loop is proportional to the small factor of $e^{-\lambda/T}$. 

\begin{figure}
\centering
\includegraphics{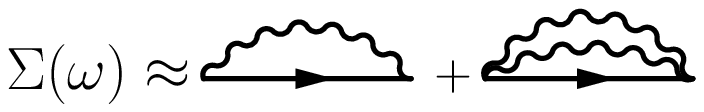}
\caption{Lower-order diagrams for pseudofermion self-energy part.
\label{sigma}} 
\end{figure}

\begin{figure}
\centering
\includegraphics{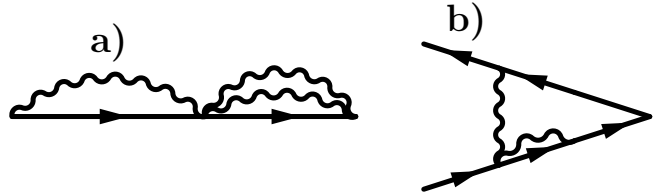}
\caption{Examples of diagrams for pseudofermion self-energy part (a) and vertex (b) containing vertexes with two and three magnons lines. Such diagrams are negligibly small compared to those presented in Figs.~\ref{sigma} and \ref{gammaf}.
\label{extra}} 
\end{figure}

Within SWA discussed in I only one virtual magnon can be emitted or absorbed in each vertex. Beyond SWA one has to consider diagrams in which some vortexes contain two and three magnons lines. Then, within SWA products containing many operators of host spins reduce to products of two-spins Green's functions while there is no such simplification beyond SWA. Meanwhile we find below that within the first order of $g^2$, the order in which all calculations of the present paper are done, diagrams with single and double wavy lines that do not contain vertexes with three wavy lines give the largest contributions. These are diagrams shown in Figs.~\ref{chifig} and \ref{sigma}. Fig.~\ref{extra} gives examples of diagrams that should be discarded. A double wavy line corresponds to longitudinal susceptibility of the host spins. Thus, the diagrammatic technique used in I can be applied for the present task with minor modification: single wavy line and double wavy line correspond to $\pi^{-1}N(\omega){\rm Im}\Delta_{\mu\nu}(\omega)$ with $\mu,\nu=x,y$ and  $\pi^{-1}N(\omega){\rm Im}\Delta_{zz}(\omega)$, respectively, where $N(\omega) = (e^{\omega/T}-1)^{-1}$ is the Planck's function,
\begin{equation}
\label{deltaf}
\Delta_{\mu\nu}(\omega) = -i\int_0^\infty dt e^{i\omega t} \langle [s_1^\mu(t) + s_2^\mu(t),s_1^\nu(0) + s_2^\nu(0)]\rangle
\end{equation}
and $\langle\dots\rangle$ denotes the thermal average; frequencies of functions ${\rm Im}\Delta_{\mu\nu}(\omega)$ should be taken so as they are contained in arguments of $G^{(0)}$-functions with positive sign; integration over all frequencies of functions ${\rm Im}\Delta_{\mu\nu}(\omega)$ is taken in the interval $(-\infty,\infty)$. ${\rm Im}\Delta_{\mu\nu}(\omega)$ is referred to throughout this paper as spectral function.

It is shown in Appendix~\ref{delprop} that the imaginary part of $\Delta_{\mu\nu}(\omega)$ given by Eq.~(\ref{deltaf}) has the following form in the ordered 2D AF at $|\omega|\gg \eta$:
\begin{eqnarray}
\label{delta}
{\rm Im}\Delta_{\mu\nu}(\omega) &=& -A\left(\frac{\omega}{\Theta}\right)^2 {\rm sgn}(\omega)\Lambda(\omega) d_{\mu\nu}
- B\frac{r(T)}{s} \frac\omega\Theta\left(\frac T\Theta\right)^2 \Lambda(\omega)\delta_{\mu z}\delta_{\nu z},\\
\label{dtensor}
d_{\mu\nu} &=& \delta_{\mu\nu} (1 - \delta_{\mu z}) + \frac{v(T)}{s}\delta_{\mu z}\delta_{\nu z},
\end{eqnarray}
where $A$ and $B$ are positive constants given by Eqs.~(\ref{a}) and (\ref{b}), respectively, that are independent of $s$ and which dimensionality is inverse energy, $\Theta$ is a characteristic energy for which we have Eq.~(\ref{theta}) and $\Lambda(\omega)$ is a cut-off function that is equal to unity at $|\omega| < \Theta$ and it drops rapidly to zero outside this interval. Within the first order of $1/s$ we have
\begin{equation}
\label{vr}
r(T) = 1, \qquad v(T) = \frac{8sJ}{N} \sum_{\bf k} \frac{N(\epsilon_{\bf k})}{\epsilon_{\bf k}},
\end{equation}
where $N$ is the number of spins in the lattice and $\epsilon_{\bf k}$ is the spin-wave energy which is equal to $\sqrt8sJk$ at small $k$. The logarithmic infra-red singularity in the expression for $v(T)$ is screened by the interaction stabilizing the long range order at $T\ne0$: $v(T) \cong T/(\pi sJ) \ln(T/(s\eta))$. Notice that only the term proportional to $A\omega^2d_{\mu\nu} (1 - \delta_{\mu z})$ was obtained in I within SWA. It originates from transverse host spins fluctuations and corresponds to emission or absorption of one spin wave by the impurity. The remaining two contributions to ${\rm Im} \Delta_{\mu\nu}(\omega)$ in Eq.~(\ref{delta}) are of the next order of $1/s$. They stem from longitudinal host spins fluctuations. The last term in Eq.~(\ref{delta}) is proportional to $T^2\omega$ and corresponds to scattering of one spin wave on the impurity. This term can be larger than $\omega^2$. The correction proportional to $v(T)\omega^2$ originates from two kinds of processes: emission or absorption of two spin waves by the impurity and scattering of one magnon on the impurity. To determine range of values of $v(T)$ we remind the well-known formula for the average $z$ component of the host spin
\begin{equation}
\label{sz}
\langle s^z \rangle = s - \frac1N \sum_{\bf k} \frac{4sJ - \epsilon_{\bf k}}{2\epsilon_{\bf k}} -\frac{4sJ}{N} \sum_{\bf k} \frac{N(\epsilon_{\bf k})}{\epsilon_{\bf k}}.
\end{equation}
The second and the third terms in Eq.~(\ref{sz}) are equal approximately to 0.2 and $T/(2\pi sJ) \ln(T/[s\eta])$, respectively. The third term must be much smaller than $s$ within spin-wave approach. Comparing it with $v(T)$ given by Eq.~(\ref{vr}) one concludes that $v(T)/(2s)$ is much smaller than unity.

As it is explained in Appendix~\ref{delprop}, higher order $1/s$-corrections give contributions proportional to products of $\omega^2$ ($\omega T^2$) and powers of $[T/(s\Theta)] \ln(T/[s\eta])$. Hence, $r(T)$ and $v(T)$ appear to be series in powers of $[T/(s\Theta)] \ln(T/[s\eta])$. We restrict ourself in this paper by the first $1/s$-correction to $r(T)$ and $v(T)$.

Term in Eq.~(\ref{delta}) proportional to $v(T)\omega^2$ will result in the temperature corrections to some constants in $\chi(\omega)$ obtained within SWA. It will give also new logarithmic contribution to $\chi_\perp(0)$. The last term in Eq.~(\ref{delta}) will lead to the large renormalization of the Lorenz peaks widths.

We show in Appendix~\ref{delprop} that one can lead to expressions (\ref{delta})--(\ref{vr}) considering isotropic 2D AF at $T\ne0$. However there is no parameter in this case to screen infra-red singularity of the function $v(T)$. This singularity signifies that host excitations with wavelengths greater than the correlation length $\xi$ play an important part in the impurity dynamics. Then spin-wave formalism is inadequate in this case and our results are applicable for ordered 2D AFs only.

\subsection{Dynamical susceptibility of the impurity}

We have for the dynamical susceptibility after the analytical continuation from the discrete frequencies to the real axis \cite{mal2,mal4,mal5,i}
\begin{eqnarray}
\label{chi}
\chi_P(\omega) &=& \frac{e^{-\lambda/T}}{2\pi i {\cal N}}
\int^\infty_{-\infty}
dx e^{-x/T}
 {\rm Tr} \{P\left[
G(x+\omega)\Gamma_P^{++}(x+\omega,x)G(x)
-
G^*(x)\Gamma_P^{--}(x,x-\omega)G^*(x-\omega)
\right.
\nonumber\\
&&-
\left.
G(x+\omega)\Gamma_P^{+-}(x+\omega,x)G^*(x)
+
G(x)\Gamma_P^{+-}(x,x-\omega)G^*(x-\omega)
\right]\},
\end{eqnarray}
where ${\cal N}$ is the number of pseudofermions that is proportional to $e^{-\lambda/T}$, $P$ is a projection of impurity spin, $G(\omega)$ is the retarded Green's function, the trace is taken over projections of the impurity spin and signs at superscript of $\Gamma_P$ denote those of imaginary parts of the corresponding arguments [e.g., $\Gamma_P^{+-}(x,y) = \Gamma_P(x+i\delta,y-i\delta)$]. An energy shift by $\lambda$ has been performed during the derivation of Eq.~(\ref{chi}). As a result the Fermi function $(e^{(x+\lambda)/T} + 1)^{-1}$ has been replaced by $e^{-(x+\lambda)/T}$ and the functions $G$ and $\Gamma_P$ no longer depend on $\lambda$. These are those functions we calculate in the next section by the diagrammatic technique. It is clear that the bare pseudofermion Green's functions in this case are $G^{(0)}_{mm'}(\omega) = \delta_{mm'}/\omega$.

We derive below analytical expressions for the dynamical susceptibility of the impurity. Perturbation theory is used for this purpose according to the interaction. It can be done if the dimensionless constants
\begin{equation}
\label{f}
f^2 = \frac{g^2A}{\Theta} \mbox{ and } h^2 = \frac{g^2Br(T)}{s\Theta}
\end{equation}
are small. Two constants $f$ and $h$ both proportional to $|g|$ are introduced to distinguish contributions from the terms in Eq.~(\ref{delta}) proportional to $\omega^2$ and $T^2\omega$. Notice that $h=0$ within SWA. Using Eqs.~(\ref{theta}), (\ref{a}) and (\ref{b}) we have $f^2 = (g/J)^2 \sqrt\pi/(4s)$ and  $h^2 = (g/J)^2 \pi^2/s^2$. Then $g$ can be even greater than $J$ at large enough $s$.

\section{Pseudofermion Green's function and the vertex}
\label{gfver}

\subsection{Green's function}
\label{pfgf}

We turn to the calculation of the pseudofermion Green's function $G_{mn}(\omega)$. The Dyson equation for it has the following form: $\omega G_{mn}(\omega) = \delta_{mn} + \sum_q\Sigma_{mq}(\omega)G_{qn}(\omega)$, where $\Sigma_{mq}(\omega)$ are matrix elements of the self-energy. It is easy to show that matrices $\Sigma(\omega)$ and $G(\omega)$ are diagonal: $\Sigma_{mn}(\omega) = \delta_{mn}\Sigma_n(\omega)$, $G_{mn}(\omega) = \delta_{mn}G_n(\omega) = \delta_{mn}/[\omega - \Sigma_n(\omega)]$. As we demonstrate in I, there is further simplification in the case of two-level impurity: $\Sigma(\omega)$ and $G(\omega)$ are proportional to the unitary matrix.

Diagrams of the order of $g^2$ for $\Sigma_n(\omega)$ are shown in Fig.~\ref{sigma}. Let us represent the Green's function in the form
\begin{equation}
\label{gf}
G_n(\omega) = \frac{1-Z_n(\omega)}{\omega + c_n + i \gamma_n(\omega+c_n)},
\end{equation}
where $Z_n(\omega)$ and $\gamma_n(\omega)$ are some functions, $\gamma_n(\omega)$ are real ones, and $c_n$ are constants. Using Eq.~(\ref{gf}) we have in the first order of $g^2$
\begin{eqnarray}
\label{z}
Z^{(1)}_n(\omega) &=& \frac{f^2 R^{(1)}_{\Sigma n}}{\pi\Theta}\int_{-\infty}^\infty dx \frac{|x|N(x)\Lambda(x)}{x+\omega+c_n+i\gamma_n(x+\omega+c_n)},\\
\label{gamm}
\gamma_n^{(1)}(\omega) &=& h^2 R_{\Sigma n}^{(2)} \left(\frac T\Theta \right)^2 \omega [1+N(\omega)]\Lambda(\omega),\\
\label{c}
c_n^{(1)} &=& 
g^2 \frac{ S(S+1) - n^2 }{\pi} \int_{-\infty}^\infty dx \frac{N(x) {\rm Im}\Delta_\perp(x)}{x} 
+
g^2 \frac{n^2}{\pi} \int_{-\infty}^\infty dx \frac{N(x) {\rm Im}\Delta_{zz}(x)}{x}
\end{eqnarray}
where $\Delta_\perp(\omega) = \Delta_{xx}(\omega) = \Delta_{yy}(\omega)$, the principal value of the integrals is implied in Eq.~(\ref{c}) and
\begin{eqnarray}
\label{r1}
R^{(1)}_{\Sigma n} &=& (S^\mu S^\nu)_{nn} d_{\mu\nu} = S(S+1) - \left(1 - \frac{v(t)}{s}\right)n^2,\\
R^{(2)}_{\Sigma n} &=& (S^z S^z)_{nn} = n^2.
\end{eqnarray}
Summation over repeated Greek indices is understood in Eq.~(\ref{r1}) and below. The logarithmic divergence in expression (\ref{z}) at real $\omega$ is screened by the term $i\gamma_n(x+\omega+c_n)$ in the denominator. We find that higher order diagrams both in $g$ and $1/s$ (e.g., that shown in Fig.~\ref{extra} (a)) give negligibly small contributions to the self-energy part. It was obtained in I that within SWA the first nonzero contribution to $\gamma_n(\omega)$ is of the order of $f^4\propto g^4$. As is seen from Eq.~(\ref{gamm}), longitudinal host spins fluctuations lead to the correction of the order of $h^2\propto g^2$. It is important to note that $\gamma_n(\omega)$ is the constant at $|\omega| \ll T$:
\begin{equation}
\label{gamm0}
\gamma_n(\omega\to0) \approx \varGamma_{0n} = h^2 R_{\Sigma n}^{(2)} \Theta\left(\frac T\Theta \right)^3.
\end{equation}
Notice that within SWA this constant shows the same $T$-dependence: \cite{i} $\varGamma_0^{(swa)} \propto f^4 T^3/\Theta^2$.

Using Eqs.~(\ref{gf})--(\ref{c}) the number of pseudofermions $\cal N$ can be calculated up to the first order of $g^2$ with the result
\begin{equation}
\label{nc}
{\cal N} = -\sum_n\frac1\pi\int_{-\infty}^\infty dx N(x) {\rm Im} G_n(\omega)\approx e^{-\lambda/T}\sum_n e^{c_n/T}.
\end{equation}
Notice that terms proportional to $g^2$ cancel each other in the right part of Eq.~(\ref{nc}). 

It should be stressed that one can not use in Eq.~(\ref{c}) expression (\ref{delta}) for imaginary part of $\Delta_{\mu\nu}(x)$ because large $x$ are essential in the integrals. We have to use exact expressions for ${\rm Im}\Delta_{\mu\nu}(x)$ to find constants $c$: Eq.~(\ref{delr}) for ${\rm Im}\Delta_\perp(x)$,  Eq.~(\ref{perpcor}) for $1/s$-correction to ${\rm Im}\Delta_\perp(x)$, and Eq.~(\ref{deltazz}) for ${\rm Im}\Delta_{zz}(x)$. As a result one obtains up to unimportant constant independent of $n$
\begin{equation}
\label{c2}
c_n^{(1)} = - \left({\cal C} - \frac{{\cal C}_\perp + {\cal C}_\|(T)}{2s}\right) n^2 \frac{g^2}{J} = -\tilde {\cal C}(T) n^2 \frac{g^2}{J},
\end{equation}
where
\begin{eqnarray}
{\cal C} &=& \frac{J}{(2\pi)^2}\int d{\bf k} \frac{1+\cos({\bf kR}_{12})}{J_{\bf 0} + J_{\bf k}} \approx 0.25,\\
{\cal C}_\perp &=& {\cal C } \left ( 1 - \frac{1}{(2\pi)^2} \int d{\bf k} \frac{\epsilon_{\bf k}}{sJ_{\bf 0}}\right) \approx 0.04 ,\\
{\cal C}_\|(T) &=& 
\frac{sJ}{(2\pi)^4}
\int d{\bf k}_1 d{\bf k}_2 
\frac{1-\cos([{\bf k}_1 + {\bf k}_2]{\bf R}_{12})}{\epsilon_{\bf k_1} \epsilon_{\bf k_2}} 
\frac{(sJ_{\bf 0})^2 + s^2 J_{\bf k_1}J_{\bf k_2} - \epsilon_{\bf k_1} \epsilon_{\bf k_2}}{\epsilon_{\bf k_1} + \epsilon_{\bf k_2}} 
+
4 {\cal C} \left(
\frac{4sJ}{(2\pi)^2} \int d{\bf k} \frac{N(\epsilon_{\bf k})}{\epsilon_{\bf k}}
\right)
\nonumber\\
& \approx & 0.13 + \frac{4sJ}{(2\pi)^2} \int d{\bf k} \frac{N(\epsilon_{\bf k})}{\epsilon_{\bf k}},
\end{eqnarray}
where ${\bf R}_{12}$ is a vector connected host spins coupled to the impurity, $J_{\bf k} = 2J(\cos k_x + \cos k_y)$, $\epsilon_{\bf k} = s\sqrt{J_{\bf 0}^2-J_{\bf k}^2}$ is the spin-wave energy and integrals are over the chemical Brillouin zone. The constant $\cal C$ is of the zeroth order of $1/s$. It originates from transverse host spins fluctuations. Constants ${\cal C}_\perp$ and  ${\cal C}_\|(T)$ are first $1/s$-corrections stemming from transverse and longitudinal fluctuations, respectively. It is seen that their sum is of the order of $\cal C$ at small $s$ and it should be taken into account. Temperature corrections are considered in ${\cal C}_\|(T)$ only because those to ${\cal C}_\perp$ are much smaller. We note that the thermal correction to ${\cal C}_\|(T)$ coincide with that to $\langle s^z\rangle$ given by Eq.~(\ref{sz}). It should be pointed out that this correction reduces the value of $\tilde {\cal C}(T)$ and can change its sign at $T\alt T_N$.
\footnote{This consideration is qualitative as spin-wave formalism works badly at such high temperatures.}

Let us discuss the physical meaning of constants $c$ given by Eq.~(\ref{c2}). They describe splitting of the impurity caused by the host spins fluctuations. The distances between the levels are given by differences between corresponding constants $c$. One infers from Eq.~(\ref{nc}) that at $T$ much smaller than the distance between the nearest levels only those with maximum $c$ are populated. As is clear from Eq.~(\ref{c2}), these are levels with the smallest in absolute value projections on the quantized axis. Then, the impurity splitting can be described by the effective Hamiltonian $- S_z^2 \tilde {\cal C}(T)g^2/J$ that is the effective one-ion anisotropy. Such anisotropy was obtained within SWA for frustrated defects in 3D AF. \cite{shender,ivanov} 

It is also seen from Eq.~(\ref{c2}) that spin-$\frac12$ defect remains degenerate because $c_{\frac12}=c_{-\frac12}$. We have taken advantage of this circumstance in I and discarded constants $c$ attributing them to renormalization of the chemical potential $\lambda$.

\subsection{Pseudofermion vertex}
\label{vertex}

\begin{figure}
\centering
\includegraphics{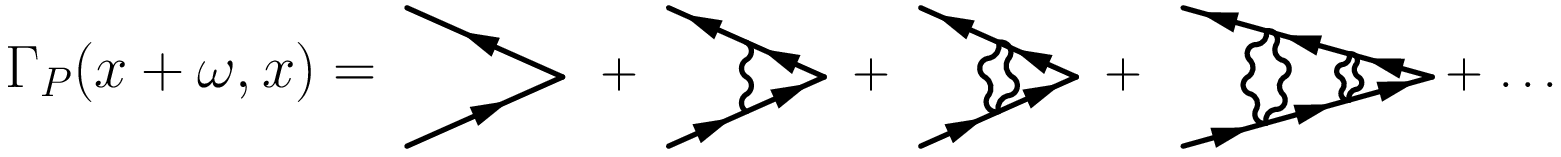}
\caption{Lower-order diagrams for pseudofermion vertex $\Gamma_P (x+\omega,x)$.
\label{gammaf}} 
\end{figure}

Let us turn to the consideration of the pseudofermion vertex $\Gamma_P(x+\omega,x)$. First diagrams for this quantity are presented in Fig.~\ref{gammaf}. It is shown in I that $\Gamma_{Pmm'}(x+\omega,x)$ is proportional to $P_{mm'}$ for $S=1/2$. However in general it has a more complicated matrix structure.

As is seen from Eq.~(\ref{chi}), we need four different branches of $\Gamma_P(x+\omega,x)$. It is clear that $\Gamma^{++} = (\Gamma^{--})^*$ and within the first order of $g^2$ one has
\begin{equation}
\label{g++}
\Gamma_{Pmn}^{++}(x+\omega,x) = P_{mn} + \frac{f^2}{\pi\Theta} \sum_{l,q} S^\nu_{ml} P_{lq} S^\mu_{qn} d_{\nu\mu} \int_{-\infty}^\infty dy y|y|N(y)\Lambda(y) G_l(x+y+\omega)G_q(x+y).
\end{equation}
It is seen from Eq.~(\ref{g++}) that the poles of $G$-functions are on the one hand from the real axis. Hence, the second term in Eq.~(\ref{g++}) is small compared to the first one and we can restrict ourselves by this precision. The term proportional to $h^2$ is discarded in Eq.~(\ref{g++}) being negligibly small due to the factor $(T/\Theta)^2$. 

The situation is different in the case of $\Gamma^{+-} = (\Gamma^{-+})^*$. Poles of the Green's functions under integrals appear to be on the opposite sides of the real axis. As a result at $\omega=0$ the integral diverges at finite $x$ as widths $\varGamma_0$ and constants $c$ tend to zero. Then one has to consider entire series to find $\Gamma^{+-}$. After analysis of diagrams for $\Gamma^{+-}$ and evaluation of their contribution to $\chi_P(\omega)$ we find that the most important diagrams within each order of $g^2$ are taken into account in the following equation:
\begin{eqnarray}
\label{g+-}
\Gamma_{Pmn}^{+-}(x+\omega,x) &=& 
P_{mn} + 
\frac{f^2}{\pi\Theta} \sum_{l,q} S^\nu_{ml} P_{lq} S^\mu_{qn} d_{\nu\mu} \int_{-\infty}^\infty dy y|y|N(y)\Lambda(y) G_l(x+y+\omega)G^*_q(x+y)
\nonumber\\
&&{}+
\frac{h^2T^2}{\pi\Theta^2} S^z_{mm}S^z_{nn} \int_{-\infty}^\infty dy y N(y)\Lambda(y) \Gamma_{Pmn}^{+-}(x+y+\omega,x+y) G_m(x+y+\omega)G_n^*(x+y).
\end{eqnarray}
Notice that this equation differs from that obtained in I within SWA. It takes into account only ladder diagrams shown in Fig.~\ref{gammaf} whereas within SWA one has to consider also diagrams with crossing of two neighboring rungs. Eq.~(\ref{g+-}) can be easily solved if one notes that the last term in this equation should be taken into account only when $|\omega + c_m - c_n|\ll T$. Outside this interval the last term is much smaller than the second one. At such $\omega$ the area of integration near the poles of the Green's functions is essential in the last term in Eq.~(\ref{g+-}) and it gives the resonant contribution:
\begin{eqnarray}
\label{eq}
\Gamma_{Pmn}^{+-}(x+\omega,x) &=&  
P_{mn} + 
\frac{f^2}{\pi\Theta} \sum_{l,q} S^\nu_{ml} P_{lq} S^\mu_{qn} d_{\nu\mu} \int_{-\infty}^\infty dy y|y|N(y)\Lambda(y) G_l(x+y+\omega)G^*_q(x+y)
\nonumber\\
&&{}-
2ih^2 S^z_{mm}S^z_{nn} \left(\frac T\Theta\right)^2 (x+c_n)N(-x-c_n) \Lambda(x+c_n)\frac{ \Gamma_{Pmn}^{+-}(\omega-c_n,-c_n)}{\omega + c_m - c_n + i\varGamma_{0m} + i\varGamma_{0n}}.
\end{eqnarray}
Eq.~(\ref{eq}) has the same structure as the equation derived in I. The difference is that $\varGamma_0$ and the third term are of the order of $g^2$ rather than $g^4$. Solving Eq.~(\ref{eq}) one obtains
\begin{eqnarray}
\label{gam+-}
\Gamma_{Pmn}^{+-}(x+\omega,x) &=& 
P_{mn} + 
\frac{f^2}{\pi\Theta} \sum_{l,q} S^\nu_{ml} P_{lq} S^\mu_{qn} d_{\nu\mu} \int_{-\infty}^\infty dy y|y|N(y)\Lambda(y) G_l(x+y+\omega)G^*_q(x+y)
\nonumber\\
&&{}+
2ih^2 S^z_{mm}S^z_{nn}P_{mn} \Theta\left(\frac T\Theta\right)^3 \frac{ \Lambda(x+c_n) }{\omega + c_m - c_n + 2i\varGamma_{mn}},\\
\varGamma_{mn} &=& h^2 (m-n)^2 \frac\Theta2 \left(\frac T\Theta \right)^3.
\end{eqnarray}
We use in Eq.~(\ref{gam+-}) that the following $x$ will be important for calculation contributions to $\chi_P(\omega)$ from $\Gamma^{+-}$: $|x+c_n|\ll T$. Note, the third term in Eq.~(\ref{gam+-}) is much greater than the second one when $|\omega+c_m-c_n|\ll \varGamma_0/f^2$. It should be pointed out that $\varGamma_{mn}=0$ for $P=S^z$. Evidently, temperature dependences of $\varGamma_{mn}$ for $P = S^x,S^y$ and $\varGamma_0$ given by Eq.~(\ref{gamm0}) are the same. Within SWA the constant $\varGamma_{mn}$ has the same $T$-dependence but it is of the order of $g^4$: \cite{i} $\varGamma_{mn}^{(swa)} \propto f^4 T^3/\Theta^2$.

\section{Impurity susceptibility}
\label{ds}

We can derive now the impurity susceptibility using the general expression (\ref{chi}), Eqs.~(\ref{gf}), (\ref{z}) and (\ref{gamm}) for the Green's function, Eqs.~(\ref{g++}) and (\ref{gam+-}) for the branches of the vertex and Eq.~(\ref{nc}) for the number of pseudofermions $\cal N$. It is convenient to discuss separately the cases of $S=1/2$ and $S>1/2$.

\subsection{Spin-1/2 impurity}

\begin{figure}
\centering
\includegraphics[scale=0.8]{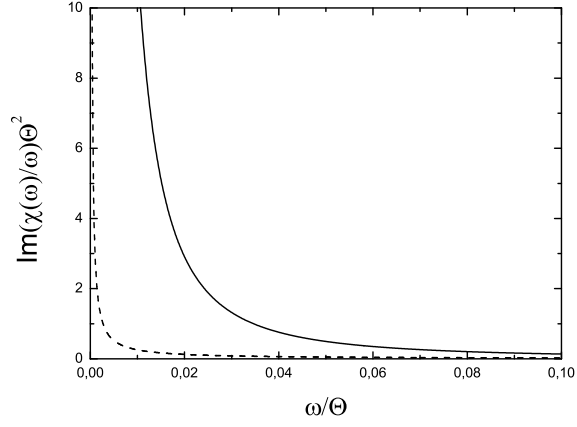}
\caption{The even-$\omega$ function ${\rm Im} (\chi_\perp(\omega)/\omega)\Theta^2$ for spin-$\frac12$ impurity at $T/\Theta=0.1$ and $f=0.1$ with longitudinal host spins fluctuations being (solid line) and not being (dotted line) taken into account.
\label{chi12gr}} 
\end{figure}

As a result of tedious calculations some details of which are presented in Appendix~\ref{gameval} we have for the dynamical susceptibility of the impurity up to terms of the order of $g^2$
\begin{eqnarray}
\label{chi2}
\chi_P(\omega) &=& \frac{1}{4T}
\frac{2i\varGamma}{\omega+2i\varGamma}
+
2R_\chi\frac{f^2}{\pi\Theta} \int_{-\infty}^\infty dx \frac{{\rm sgn}(x)\Lambda(x)}{x + \omega + 2i\varGamma_0},\\
\label{rchi}
R_\chi &=& \frac{\overline{PS^\mu[S^\nu, P]}d_{\mu\nu}}{2S+1},
\end{eqnarray}
where we introduce the notation $\overline Y = {\rm Tr}(Y)$, $\varGamma_0=h^2T^3/(4\Theta^2)$, $\varGamma=0$ for $P=S^z$ and $\varGamma=h^2T^3/(2\Theta^2)$ for $P=S^{x,y}$. Expression (\ref{chi2}) differs from that obtained in I only by the form of $\varGamma_0$ and $\varGamma$ and by the thermal $1/s$-correction to $R_\chi$. The first term in Eq.~(\ref{chi2}) is the Lorenz peak with the width $\varGamma$. The last term is the non-resonant part of the susceptibility. Its imaginary part at $|\omega|\gg\varGamma_0$ is proportional to ${\rm sgn}(\omega)$ and the real one contains the logarithmic singularity of the form $\ln (\omega^2 + \varGamma_0^2)$. At $T=0$ and $\omega\ne0$ the nonresonant contribution survives only and the susceptibility has the logarithmic singularity. It is clear that longitudinal host spins fluctuations become important if $\varGamma_0 \gg \varGamma_0^{(swa)}$ and $\varGamma \gg \varGamma^{(swa)}$. These conditions are fulfilled when $|g|/J \ll 5$, i.e., for weakly coupled defects being considered. The contrast between results which do and do not take into account the longitudinal host spins fluctuations is illustrated by Fig.~\ref{chi12gr}, where we plot ${\rm Im}\chi_\perp(\omega)/\omega$ using Eq.~(\ref{chi2}).

The non-resonant term gives the main contribution to the susceptibility (\ref{chi2}) when
\begin{equation}
\label{omlim}
\omega \gg \omega_0 = \frac \Theta s \left( \frac T\Theta \right)^2.
\end{equation}
Notice that $\omega_0\propto f^2T^2$ within SWA. 

Using results of calculations presented in Appendix~\ref{gameval} we find that static susceptibility $\chi_P(0)$ does not depend on $\varGamma$ and $\varGamma_0$:
\begin{eqnarray}
\label{w2}
\chi_P(0) &=& \frac{1}{4T} \left(1 - 4f^2{\cal U}R_\chi \right) 
+
4 R_\chi\frac{f^2}{\pi\Theta}\ln\left(\frac \Theta T\right),\\
\label{uchi}
{\cal U} &=& -\frac{4SJ^2}{\pi^{3/2}}\int dx \frac{{\rm Im} \Delta_\perp(x)}{x|x|} 
= 
\frac{2J^2}{\pi^{5/2}}\int d{\bf k} \frac{[1+\cos({\bf kR})][J_{\bf 0} - J_{\bf k}]}{(J_{\bf 0}^2 - J_{\bf k}^2)^{3/2}}
\approx 0.39.
\end{eqnarray}
It is taken into account in Eq.~(\ref{uchi}) that one cannot use Eq.~(\ref{delta}) for ${\rm Im}\Delta_\perp(x)$ because large $x$ are important in the integral. 
\footnote{This circumstance has not been taken into account in I where $f^2$-correction to $1/T$-term was calculated incorrectly.} 
Eq.~(\ref{delr}) is used to find $\cal U$ as it was done above for calculation of constants $c$. It is seen from Eq.~(\ref{w2}) that the static susceptibility contains the free-spin-like term $(4T)^{-1}$ which amplitude is slightly reduced by the interaction and the logarithmic correction proportional to $f^2$. Static susceptibility has the same form as that obtained in I within SWA with the only difference in the form of the constant $R_\chi = R_\chi^{(swa)} + [v(T)/(2s)]\overline{PS^z[S^z, P]}$. Notice that $R_\chi = R_\chi^{(swa)} = 1/4$ for $P=S^z$ and $\chi_\|(0) = \chi_\|^{(swa)}(0)$. In contrast $R_\chi = R_\chi^{(swa)} + v(T)/(8s) = 1/8 + v(T)/(8s)$ for $P=S^{x,y}$ and the transverse static susceptibility has the form
\begin{equation}
\label{stperp}
\chi_\perp(0) = \frac{1}{4T} \left(1 - f^2\frac{\cal U}{2} \right) 
-
f^2\frac{\cal U}{\sqrt\pi s\Theta}\ln\left(\frac{T}{s\eta}\right)
+
\frac{f^2}{2\pi\Theta}\ln\left(\frac \Theta T\right).
\end{equation}
Thus the longitudinal host spins fluctuations lead to another logarithmic term in $\chi_\perp(0)$.

\subsection{Impurities with $S > 1/2$}

\subsubsection{Dynamical susceptibility}

The calculations are slightly more tedious for impurity with $S>1/2$. As a result of simple evaluations presented in Appendix~\ref{gameval} we lead to the following quite a cumbersome expression for the impurity susceptibility: 
\begin{eqnarray}
\label{chi-l}
\chi(\omega) &=& 
\frac{e^{-\lambda/T}}{\cal N}
\sum_{m,n}
\left[
|P_{nm}|^2
\left(
e^{(c_m+2i\varGamma_{nm}-i\varGamma_{0n})/T} - e^{(c_n-i\varGamma_{0n})/T}
\right)
\frac{1}{\omega + c_m - c_n + 2i\varGamma_{nm}}
\right.
\nonumber\\
&&{}+
|P_{nm}|^2
\left(R_{\Sigma n}^{(1)} + R_{\Sigma m}^{(1)}\right) \left(e^{c_n/T} + e^{c_m/T}\right) \frac{f^2}{2\pi\Theta}\int_{-\infty}^\infty dx \frac{{\rm sgn}(x)\Lambda(x)}{x+\omega + c_m - c_n + i\varGamma_{0n} + i\varGamma_{0m}}
\nonumber\\
&&{}+
\frac{2{\rm Re}(P_{nm}S_{mq}^\nu P_{ql} S_{ln}^\mu d_{\mu\nu})}{\omega + c_m - c_n + i\varGamma_{0n} + i\varGamma_{0m}}\frac{f^2}{\pi\Theta}
\int_{-\infty}^\infty dx |x|xN(x)\Lambda(x) 
\nonumber\\
&&\times{}
\left(
\frac{e^{(c_m+i\varGamma_{om})/T}}{ (x-\omega+c_l-c_m-i\varGamma_{0l}-i\varGamma_{0m}) (x+c_q-c_m+i\varGamma_{0q}-i\varGamma_{0m})}
\right.\nonumber\\
&&{}-
\left.
\left.
\frac{e^{(c_n-i\varGamma_{on})/T}}{ (x+\omega+c_q-c_n+i\varGamma_{0q}+i\varGamma_{0n}) (x+c_l-c_n+i\varGamma_{0n}-i\varGamma_{0l})}
\right)
\right].
\end{eqnarray}
The first term here is zero for longitudinal susceptibility ($P=S^z$ and thus $m=n$). In the case of transverse susceptibility it gives $2S$ Lorenz peaks corresponding to transitions between impurity levels. Using Eq.~(\ref{c2}) we obtain for the resonant frequencies
\begin{equation}
\label{ores}
\omega^{(n)}_{res} = |c_n-c_{n-1}| = (2n-1) \tilde {\cal C}(T) g^2/J, 
\end{equation}
where $n=S,S-1,\dots,n_{\rm min}$ and $n_{\rm min}=1/2$ and 1 for half-integer and integer spins, respectively. Then in contrast to half-integer spins there is no resonance peak at zero frequency in the case of integer spins. Notice that the distance between frequencies does not depend on $S$. The second and the third terms in Eq.~(\ref{chi-l}) give contributions to the non-resonant term. The third term contains also a resonant contribution. As is explained in Appendix~\ref{gameval}, this contribution should be discarded because the resonant terms of the susceptibility are calculated within the zeroth order of $g^2$.

One makes sure after simple transformations that Eq.~(\ref{chi-l}) coincides with Eq.~(\ref{chi2}) if $S=1/2$. It is more convenient to discuss expression (\ref{chi-l}) further in some limiting cases in which $\chi(\omega)$ has simpler forms. Let us consider the susceptibility when temperature is greater than the impurity splitting, $T\gg \tilde {\cal C}(T) g^2S^2/J$, and when $T$ is much smaller than the energy between nearest impurity levels, $T\ll \tilde {\cal C}(T)g^2/J$.

\begin{figure}
\centering
\includegraphics[scale=0.8]{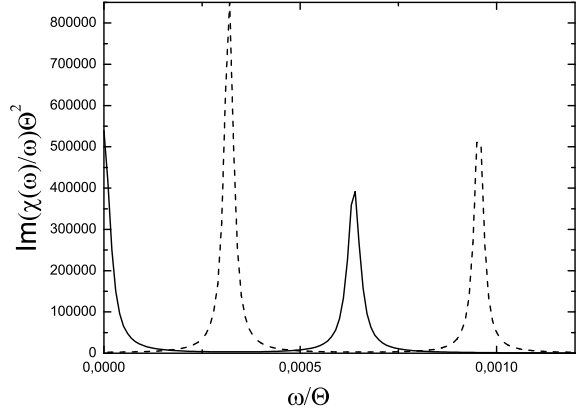}
\caption{The even-$\omega$ function ${\rm Im} (\chi_\perp(\omega)/\omega)\Theta^2$ calculated using Eq.~(\ref{chi_l>}) for impurity with $S=3/2$ (solid line) and $S=2$ (dotted line) at $T/\Theta=0.05$ and $f=0.1$.
\label{chi32gr}} 
\end{figure}

$T\gg \tilde {\cal C}(T)g^2S^2/J$. Eq.~(\ref{chi-l}) can be simplified greatly in this case if one expands exponents $e^{c_n/T}$ up to the second term and notes that the nonresonant terms should be taken into account at $|\omega|\gg \tilde {\cal C}(T) g^2S/J$ only. As a result one obtains
\begin{eqnarray}
\label{chi_l>}
\chi_P(\omega) &=& \frac{1}{T(2S+1)}
\sum_{n,m}
|P_{nm}|^2
\frac{c_m - c_n + 2i\varGamma_{nm}}{\omega + c_m - c_n + 2i\varGamma_{nm}}
+
2R_\chi\frac{f^2}{\pi\Theta}\int_{-\infty}^\infty dx \frac{{\rm sgn}(x)\Lambda(x)}{x+\omega + i\delta},
\end{eqnarray}
where $R_\chi$ is given by Eq.~(\ref{rchi}). Notice that Eq.~(\ref{chi_l>}) coincides with Eq.~(\ref{chi2}) if $S=1/2$. Imaginary part of the nonresonant term is greater than that of resonant terms when $|\omega|\gg \omega_0$, where $\omega_0$ is given by Eq.~(\ref{omlim}). The real part of the nonresonant term dominates in Eq.~(\ref{chi_l>}) at $|\omega|\gg f\Theta\sqrt{\Theta/T}$.
To illustrate our results we plot in Fig.~\ref{chi32gr} the even-$\omega$ function ${\rm Im}\chi_\perp(\omega)/\omega$ using Eq.~(\ref{chi_l>}) for $S=3/2$ and $S=2$.

$T\ll \tilde {\cal C}(T)g^2/J$. At such small temperatures only low-lying impurity levels contribute to susceptibility. The transverse susceptibility for integer impurity spin has the form
\begin{eqnarray}
\label{chi_int<}
\chi_\perp^{int}(\omega) &=& 
\frac{2|P_{01}|^2}{\omega + c_0 - c_1 + 2i\varGamma_{01}}
-
\frac{2|P_{01}|^2}{\omega - (c_0 - c_1) + 2i\varGamma_{01}}
+ i\frac{f^2}{\Theta}
\left\{
|P_{01}|^2 \left(R_{\Sigma0}^{(1)} + R_{\Sigma 1}^{(1)} \right) {\rm sgn}(\omega + c_1 - c_0)
\right.\nonumber\\
&&{}-
\left.
\sum_{m,q,l} \frac{{\rm Re}(P_{0m}S_{mq}^\nu P_{ql}S_{l0}^\mu d_{\mu\nu})}{\omega + c_m - c_0 + i\varGamma_{0m}}
\frac{(\omega + c_q - c_0)^2}{\omega + c_q - c_l + i\varGamma_{0q} + i\varGamma_{0l}}(1 + {\rm sgn}(\omega + c_q - c_0)) - \{\omega\to-\omega\}
\right\},
\end{eqnarray}
where $\{\omega\to-\omega\}$ denotes terms inside the curly brackets with $-\omega$ put instead $\omega$. The first two terms in Eq.~(\ref{chi_int<}) describe transitions between three low-lying impurity levels: $\{-1\leftrightarrow 0\}$ and $\{0\leftrightarrow 1\}$. The last term in Eq.~(\ref{chi_int<}) is the non-resonant part of the susceptibility that is zero between the resonance peaks $\omega=\pm(c_0-c_1)$. Beyond this interval it is negligibly small near the resonance peaks, i.e.\ at $|\omega + c_0 - c_1|\ll T^{3/2}/\sqrt\Theta$ and  $|\omega - (c_0 - c_1)|\ll T^{3/2}/\sqrt\Theta$, but it gives the main contribution to imaginary part of the susceptibility outside these two intervals being of the order of $f^2$. The real part of the non-resonant term is small compared to that of the resonant terms at $|\omega|<\Theta$.

In the case of half-integer impurity spins the transverse susceptibility has the form
\begin{eqnarray}
\label{chi_hint<}
\chi_\perp^{half-int}(\omega) &=& 
\frac{|P_{-\frac12\frac12}|^2}{T}\frac{2i\varGamma_{-\frac12\frac12}}{\omega + 2i\varGamma_{-\frac12\frac12}}
+
\frac{|P_{\frac12\frac32}|^2}{\omega + c_\frac12 - c_\frac32 + 2i\varGamma_{\frac12\frac32}}
-
\frac{|P_{\frac12\frac32}|^2}{\omega - (c_\frac12 - c_\frac32) + 2i\varGamma_{\frac12\frac32}}\nonumber\\
&&{} + i\frac{f^2}{4\Theta}\sum_{n=\pm\frac12}
\left[
\sum_m |P_{nm}|^2 \left(R_{\Sigma n}^{(1)} + R_{\Sigma m}^{(1)} \right) {\rm sgn}(\omega + c_m - c_n)
\right.\nonumber\\
&&{}-
\left.
\sum_{m,q,l} \frac{2{\rm Re}(P_{nm}S_{mq}^\nu P_{ql}S_{ln}^\mu d_{\mu\nu})}{\omega + c_m - c_n + i\varGamma_{0m} + i\varGamma_{0n}}
\frac{(\omega + c_q - c_n)^2}{\omega + c_q - c_l + i\varGamma_{0q} + i\varGamma_{0l}}(1 + {\rm sgn}(\omega + c_q - c_n)) - \{\omega\to-\omega\}
\right],
\end{eqnarray}
where the first three terms describe transitions between four low-lying impurity levels: $\{-3/2\leftrightarrow -1/2\}$, $\{-1/2\leftrightarrow 1/2\}$, and $\{1/2\leftrightarrow 3/2\}$. The non-resonant part of the susceptibility is negligibly small in the vicinity of the resonance peaks, i.e.\ at $|\omega|\ll T^2/\Theta$, $|\omega + c_{1/2} - c_{3/2}|\ll T^{3/2}/\sqrt\Theta$ and  $|\omega - (c_{1/2} - c_{3/2})|\ll T^{3/2}/\sqrt\Theta$. Outside these three intervals the nonresonant term is of the order of $f^2$ and it gives the main contribution to imaginary part of the susceptibility. The real part of the nonresonant term is small compared to that of the resonant terms at $|\omega|<\Theta$.

\subsubsection{Static susceptibility}

$T\gg \tilde {\cal C}(T)g^2S^2/J$. Using results of Appendix~\ref{gameval} one finds for static susceptibility in this regime
\begin{equation}
\label{chi0-l-g}
\chi_P(0) = \frac{S(S+1)}{3T}
\left(
1 - f^2 {\cal U} \frac{3R_\chi}{S(S+1)}
\right)
+
4R_\chi \frac{f^2}{\pi\Theta} \ln\left(\frac\Theta T\right),
\end{equation}
where $R_\chi$ and $\cal U$ are given by Eqs.~(\ref{rchi}) and (\ref{uchi}), respectively. Eq.~(\ref{chi0-l-g}) coincides with Eq.~(\ref{w2}) at $S=1/2$. It is seen that static susceptibility of large-$S$ impurities has the same structure as that of spin-$\frac12$ defect.

$T\ll \tilde {\cal C}(T)g^2/J$. At small temperature integer and half-integer impurity spins behave differently. Thus, transverse static susceptibility of an integer spin is a constant:
\begin{equation}
\label{chi0-l-l-i}
\chi^{int}_\perp(0) = \frac{4\left|P_{01}\right|^2}{c_0 - c_1},
\end{equation}
whereas longitudinal one is exponentially small: $\chi^{int}_\|(0) \approx 0$. 

Transverse static susceptibility of a half-integer spin has the form
\begin{eqnarray}
\label{chi0-l-l-hi}
\chi^{half-int}_\perp(0) &=& \frac1T
\left(
\left|P_{-\frac12\frac12}\right|^2 - f^2 {\cal U} Q 
\right)
+ 4Q \frac{f^2}{\pi\Theta} \ln\left(\frac\Theta T\right)
+
\frac{2\left|P_{\frac12\frac32}\right|^2}{c_{\frac12} - c_{\frac32}},\\
\label{q}
Q &=& \frac12 \left( {\sum_{n,m}}' |P_{nm}|^2 R_{\Sigma n}^{(1)} - {\sum_{n,m,q,l}}' P_{nm}S^\mu_{mq}P_{ql}S^\nu_{ln}d_{\mu\nu} \right),
\end{eqnarray}
where primes over sums signify that summation is taken over projections $1/2$ and $-1/2$ only. It is seen that to the first two terms in Eq.~(\ref{chi0-l-l-hi}) contribute only transitions between states with projections $\pm1/2$. The last term represents the contribution to the susceptibility from transitions between states with projections $\pm1/2$ and $\pm3/2$. It is of the order of $1/f^2$ being much larger than the logarithmic term up to exponentially small temperatures $T\sim \Theta e^{-1/f^4}$.

Longitudinal static susceptibility of a half-integer spin behaves like that of spin-$\frac12$ defect:
\begin{eqnarray}
\label{chi0-l-t-hi}
\chi^{half-int}_\|(0) &=& \frac1T
\left(
\left|P_{-\frac12\frac12}\right|^2 - f^2 {\cal U} Q' 
\right)
+
4Q \frac{f^2}{\pi\Theta} \ln\left(\frac\Theta T\right),\\
\label{q2}
Q' &=& \frac12 \left( {\sum_{n,m}}' |P_{nm}|^2 R_{\Sigma n}^{(1)} - \sum_{q,l}{\sum_{n,m}}' P_{nm}S^\mu_{mq}P_{ql}S^\nu_{ln}d_{\mu\nu} \right),
\end{eqnarray}
where $Q$ is given by Eq.~(\ref{q}). In contrast to $\chi^{half-int}_\perp(0)$ there is no term proportional to $1/f^2$ in $\chi^{half-int}_\|(0)$ screening the logarithmic correction.

As it is explained in I, there is a restriction on the range of validity of the resultant expressions for $\chi_P(\omega)$. It is the consequence of the fact that the function ${\rm Im}\Delta(\omega)$ has the form (\ref{delta}) if $|\omega| \gg \eta$ only. It is easy to see that in all calculations performed above one can use the function of the form (\ref{delta}) if the following condition on $\omega$ and widths $\varGamma_0$ holds: $\max\{\min\{\varGamma_{0n}\},|\omega|\} \gg \eta$.

\section{Influence of defects on the host system}
\label{influence}

We discuss in this section the influence of finite concentration of the defects $n$ on the spin-wave spectrum of AF. It is demonstrated in I that in the vicinity of the points $k = 0$ and $k = k_0$, where ${\bf k}_0$ is the antiferromagnetic vector, denominator of host spins Green's functions determined the spin-wave spectrum has the form
\begin{eqnarray}
\label{fg2}
{\cal D}(\omega,{\bf k}) &=& \omega^2 - \epsilon_{\bf k}^2 \left[ 1 - \frac{n f^2}{2\pi} \Theta u({\bf k})\chi_\perp(\omega) \right],\\
\label{u}
u({\bf k}) &=& \frac 12 + \frac{({\bf kR}_{12})^2}{k^2},
\end{eqnarray}
where ${\bf R}_{12}$ is the vector connected host spins coupled to impurity (it is assumed for beginning that this vector is the same for all defects) and it is used that the unperturbed spectrum is linear at $k\ll k_0$ and $k\sim k_0$: $\epsilon_{\bf k}=ck=\sqrt8 sJk$. It is seen from Eqs.~(\ref{fg2}) and (\ref{u}) that the spectrum depends on the direction of the momentum $\bf k$ as a result of interaction of magnons with the defects. This circumstance is a consequence of our assumption that the vector ${\bf R}_{12}$ is the same for all impurities. In fact, it can have four directions and the value $({\bf R}_{12}{\bf k})^2/k^2$ can have two different values: $\cos^2 \phi_{\bf k}$ and  $\sin^2 \phi_{\bf k}$, where $\phi_{\bf k}$ is the azimuthal angle of $\bf k$. It easy to realize that $u({\bf k})=1$ if all four ways of coupling of the impurity with AF are equally possible.

Let us turn to spin-$\frac12$ impurity. It is convenient to consider separately regimes of $|\omega| \gg \omega_0$ and $|\omega| \ll \omega_0$, where $\omega_0$ is given by Eq.~(\ref{omlim}). In these cases the non-resonant and the resonant parts, respectively, dominate in the impurity susceptibility (\ref{chi2}). The results are summarized in Table~\ref{specren}. They are very similar to those derived in I within SWA. The only difference is in the form of $\omega_0$, $\varGamma$ and $R_\chi$. It is seen that due to interaction with defects the spin-wave velocity acquires negative corrections and strong spin-wave damping arises proportional to $nf^4\omega$ at $\omega\gg \omega_0$.

As it is explained in I the results obtained within our method have a limited range of validity which is also indicated in Table~\ref{specren}. This is because the interaction with defects modifies host spins Green's functions and the bare spectral function ${\rm Im} \Delta_{\mu\nu}^{(0)}(\omega)$ given by Eq.~(\ref{delta}) renormalizes as follows: 
\begin{eqnarray}
\label{delad1}
{\rm Im} \Delta_{\mu\nu}(\omega) &=& {\rm Im} \Delta_{\mu\nu}^{(0)}(\omega) 
-
snf^2{\cal B} {\rm Im}\chi_\perp(\omega) d_{\mu\nu},\\
\label{a2}
{\cal B} &=& 
\frac{4J^2}{\pi^{5/2}}
\int  
d{\bf k}
\left(
\frac{1+\cos({\bf k R}_{12})}{J_{\bf 0} + J_{\bf k}}
\right)^2
\approx 0.8.
\end{eqnarray}
The last term in Eq.~(\ref{delad1}) becomes larger than ${\rm Im} \Delta_{\mu\nu}^{(0)}(\omega)$ at small enough energies indicated in Table~\ref{specren} and the problem should be solved self-consistently. Corresponding consideration is out of the scope of the present paper. 

This large renormalization of the spectral function at small energies can be interpreted as a resonance scattering of spin waves on impurities. Similar situation exists in the case of asymmetrically coupled defects. Let us discuss, for example, AF with an impurity spin weakly coupled to one host spin. According to results obtained within $T$-matrix approach the defect leads to a resonant level inside the spin-wave band. \cite{lov1,lov2,iz} The energy of this level is of the order of the energy cost for creation of spin excitation on the impurity. First perturbation corrections to observables increase rapidly near the resonant level due to the resonance scattering. \cite{iz} In the case of frustrated spin-$\frac12$ defect the energy of creation of the spin excitation on the impurity is zero and the resonance scattering takes place at zero energy.

In the vicinity of the resonance peak, where our theory is not applicable, one might expect strong deviation from linearity of the real part of the spin-wave spectrum and an enhancement of the spin-wave damping as in the case of asymmetrically coupled defects \cite{iz}.

Particular expressions for spin-wave velocity and damping in the case of $S>1/2$ can be obtained straightforwardly using Eqs.~(\ref{chi_l>}), (\ref{chi_int<}), (\ref{chi_hint<}) and (\ref{fg2}). Because of their cumbersomeness we do not present here the results. We would like to point out only that there is the abnormal spin-wave damping proportional to $nf^4|\omega|$ (stemming from nonresonant part of $\chi_\perp(\omega)$) when i) $|\omega|\gg T^2$ if $T\gg \tilde {\cal C}(T) g^2S^2/J$, ii) $\left||\omega| - \omega_{res}^{(n)}\right|\gg T^{3/2}$ for $n\ne1/2$ and $\left||\omega| - \omega_{res}^{(1/2)}\right|\gg T^2$ if $T\ll \tilde {\cal C}(T)g^2/J$, where $\omega_{res}^{(n)}$ are given by Eq.~(\ref{ores}). The range of validity of the results obtained within our approach can be found comparing the last term in Eq.~(\ref{delad1}) with the first one as it was done for $S=1/2$. The results are not valid in the vicinity of the resonance peaks due to the resonance scattering of spin waves on impurities. At such energies one might expect strong deviation from linearity of the real part of the spin-wave spectrum and increase of the spin-wave damping.

Thermodynamic quantities of AF with defects can be found, in principle, using impurity susceptibility. But due to the limiting range of validity of our results and due to the fact that thermodynamic quantities are expressed via integrals containing susceptibility one cannot calculate them because it is impossible to perform the integration over the essential energy area (in particular, over small energies). We explain this situation in detail in I by the example of the specific heat. We do not obtain any noticeable corrections to the specific heat after integration over energies at which our results are valid.

\section{Conclusion}
\label{concl}

In the present paper we discuss dynamical properties of an impurity spin coupled symmetrically to sublattices of 2D Heisenberg quantum antiferromagnets (see Fig.~\ref{pic}) at $T\ge0$ (existence of a small interaction stabilizing the long range order at $T\ne0$ is implied). We continue our study on this subject started in I where spin-$\frac12$ defect was discussed and the host spins fluctuations are considered within the spin-wave approximation (SWA). In the present paper we i) go beyond SWA and ii) study impurities with spins $S \ge 1/2$. We show that the effect of the host system on the defect is completely described by the spectral function as it was within SWA. It is demonstrated that longitudinal host spins fluctuations lead to two important contributions to the spectral function that is proportional to $\omega^2$ within SWA. They have the form $r(T) T^2 \omega/s$ and $\omega^2v(T)/s$, where $s$ is the value of the host spin, $r(T)$ and $v(T)$ are series in $(T/s)\ln(T/[s\eta])$. First terms in these series are given by Eq.~(\ref{vr}). 

It is demonstrated that transverse spin-$\frac12$ impurity susceptibility $\chi_\perp^{(1/2)}(\omega)$ has the same structure as that obtained within SWA: the Lorenz peak and the non-resonant term. The difference is that the width of the peak $\varGamma$ becomes larger being proportional to $f^2J(T/J)^3$ rather than $f^4J(T/J)^3$. Besides some constants in the expression for $\chi_\perp^{(1/2)}(\omega)$ acquire thermal corrections. The width of the Lorenz peak in longitudinal impurity susceptibility is zero within our precision. The resultant expression for $\chi^{(1/2)}(\omega)$ is given by Eq.~(\ref{chi2}). We observe that longitudinal host spins fluctuations lead to a new logarithmic correction to transverse static susceptibility (\ref{stperp}) proportional to $-(f^2/[Js])\ln(T/[s\eta])$. It is demonstrated that corrections to $\varGamma$ from longitudinal host spins fluctuations dominate if $|g|/J \ll 5$.

We derive also expressions for $\chi(\omega)$ for symmetrically coupled impurities with $S>1/2$. It is well-known that host spins fluctuations lead to an effective one-ion anisotropy on the impurity site in 3D AF: $-\tilde {\cal C}(T)S_z^2g^2/J$, where $\tilde {\cal C}(T)>0$. \cite{shender,ivanov} Then defects with $S>1/2$ appear to be split and magnetization of sublattices is a hard axis for them. We obtain the same one-ion anisotropy for defects in 2D AF. In particular we observe large reduction of the value of $\tilde {\cal C}(T)$ obtained within SWA by $1/s$-corrections stemming from longitudinal host spins fluctuations. We show that thermal corrections reduce $\tilde {\cal C}(T)$ further and can change the sign of $\tilde {\cal C}(T)$ at $T\alt T_N$.

Then we find that if $T$ is greater than the value of the defect splitting, transverse dynamical susceptibility contains $2S$ Lorenz peaks corresponding to transitions between impurity levels and a nonresonant contribution. Widths of the peaks are proportional to $f^2J(T/J)^3$ as in the case of $S=1/2$. When $T$ is lower than the distance between nearest impurity levels ($\sim \tilde {\cal C}(T)g^2/J$) there are only peaks corresponding to transitions between low-lying levels. Longitudinal susceptibility has only the nonresonant term. As it was in the case of spin-$\frac12$ impurity, the imaginary part of the nonresonant term is a constant that leads to abnormal spin-wave damping proportional to $nf^4|\omega|$. At $T\gg \tilde {\cal C}(T)g^2S^2/J$ static susceptibility has the same structure as that for $S=1/2$: term $S(S+1)/(3T)$ and the logarithmic correction. If $T\ll \tilde {\cal C}(T)g^2/J$, $\chi(0)$ behaves differently for integer (Eq.~(\ref{chi0-l-l-i})) and half-integer (Eqs.~(\ref{chi0-l-l-hi}) and (\ref{chi0-l-t-hi})) spins.

Corrections to the spin-wave spectrum are obtained in the case of finite concentration of spin-$\frac12$ defects. They are summarized in Table~\ref{specren}. They have the same form as those obtained within SWA. The difference is that some constants acquire thermal corrections, $\varGamma$ has another form and the quantity $\omega_0$ separating two regimes appears to be proportional to $T^2/(sJ)$ rather than $f^2T^2/J$. In particular, we find strong spin-wave damping proportional to $nf^4|\omega|$ for all $S$ originating from the non-resonant terms of the susceptibilities. 

\begin{acknowledgments}
We are thankful to A.V.\ Lazuta for stimulating discussions. This work was supported by Russian Science Support Foundation (A.V.S.), grant of President of Russian Federation MK-4160.2006.2, RFBR (Grant Nos.\ 06-02-16702 and 05-02-19889) and Russian Programs "Quantum Macrophysics", "Strongly correlated electrons in semiconductors, metals, superconductors and magnetic materials" and "Neutron Research of Solids".
\end{acknowledgments}

\appendix

\section{Calculation of ${\rm Im} \Delta_{\mu\nu}(\omega)$}
\label{delprop}

In this appendix properties of the imaginary part of the function $\Delta_{\mu\nu}(\omega)$ are discussed general expression for which is given by Eq.~(\ref{deltaf}). We show that ${\rm Im}\Delta_{\mu\nu}(\omega)$ has the form (\ref{delta}) and find constants $A$ and $B$, the characteristic energy $\Theta$ and tensor $d_{\mu\nu}$.

One has from Eq.~(\ref{deltaf})
\begin{equation}
\label{delta0}
\Delta_{\mu\nu}(\omega) = \frac 2N \sum_{\bf k} [1 + \cos({\bf k R}_{12})] \langle s^\mu_{-\bf k} s^\nu_{\bf k} \rangle_\omega, 
\end{equation}
where $\langle \dots \rangle_\omega$ denote retarded Green's function, $N$ is the number of spins in the lattice and ${\bf R}_{12}$ is the vector connecting two host spins coupled to the defect. Thus we have to calculate the spin Green's functions $\langle s^\mu_{-\bf k} s^\nu_{\bf k} \rangle_\omega$.

It is demonstrated in I that only diagonal components of $\langle s^\mu_{-\bf k} s^\nu_{\bf k} \rangle_\omega$ are nonzero. Only $xx$- and $yy$- components are nonzero within the spin-wave approximation (SWA). At $T=0$ we have for them in the isotropic 2D AF \cite{i}
\begin{eqnarray}
\label{delr}
{\rm Im} \Delta_{\mu\nu}(\omega) &=& -d_{\mu\nu}\frac{s^2}{4\pi}
\int d{\bf k} 
\Bigl\{
[1+\cos({\bf k R}_{12})][J_{\bf 0} - J_{\bf k}]
+ 
[1 + \cos([{\bf k} + {\bf k}_0]{\bf R}_{12})][J_{\bf 0} + J_{\bf k}]
\Bigr\}
\nonumber\\
&&\times\frac{1}{\epsilon_{\bf k}}
\left[\delta(\omega-\epsilon_{\bf k}) - \delta(\omega+\epsilon_{\bf k}) \right],
\end{eqnarray}
where $d_{\mu\nu} = \delta_{\mu\nu}(1 - \delta_{\mu z})$, ${\bf k}_0$ is the antiferromagnetic vector, the lattice constant is taken to be equal to unity, $J_{\bf k} = 2J (\cos k_x + \cos k_y)$, $\epsilon_{\bf k} = s\sqrt{J_{\bf 0}^2-J_{\bf k}^2}$ is the spin-wave energy and the integral is over the {\it magnetic} Brillouin zone. If $|\omega|\ll sJ$ we have $J_{\bf k}\approx J_{\bf 0} - Jk^2$, $\cos({\bf kR}_{12})\approx 1-({\bf kR}_{12})^2/2$ and $\epsilon_{\bf k}=ck=\sqrt8 sJk$. Notice that $({\bf k}_0 {\bf R}_{12}) = \pi\mod 2\pi$ if the impurity is coupled to spins from different sublattices and both terms in the curly brackets in Eq.~(\ref{delr}) are proportional to $k^2$. Then integration in Eq.~(\ref{delr}) can be easily carried out if one takes advantage of the approximation for magnons similar to Debye one for phonons: the spectrum is assumed to be linear, $\epsilon_{\bf k}=ck$, up to cut-off momentum $k_\Theta$ defined from the equation $4\pi N = V \int_0^{k_\Theta}dkk$, where $V$ is the area of the lattice. As a result we lead to terms proportional to $d_{\mu\nu}(1-\delta_{\mu z})$ in expression (\ref{delta}) for ${\rm Im} \Delta_{\mu\nu}(\omega)$ in which 
\begin{eqnarray}
\label{theta}
\Theta &=& ck_\Theta = sJ 8\sqrt\pi,\\
\label{a}
A &=& \frac{2\pi}{J}.
\end{eqnarray}
The constant $A$ should be multiplied by 2 if the defect is coupled to four host spins (two by two from each sublattice).

It is easy to conclude that if a small interplane interaction of the value of $\eta\ll J$ is taken into account the above result for ${\rm Im}\Delta_{\mu\nu}(\omega)$ is valid when $|\omega|\gg\eta$. At the same time ${\rm Im}\Delta_{\mu\nu}(\omega)$ has another $\omega$-dependence if $|\omega|\alt\eta$. It is well established \cite{kop,chak,tyc,thur} that spin waves are well defined in paramagnetic phase of 2D AF if their wavelength is much smaller than the correlation length $\xi\propto \exp({\rm const}/T)$. Thus, the above result for ${\rm Im}\Delta_{\mu\nu}(\omega)$ is valid when $|\omega|\gg Ja/\xi$.

Let us go beyond SWA. It is well known that the spin-wave interaction gives negligibly small corrections to the spin-wave spectrum and other physical quantities in isotropic Heisenberg AFs. \cite{kop,cast,canali,igar2,harris,malold} For instance, simple calculations give that transverse components of ${\rm Im}\Delta_{\mu\nu}(\omega)$ given by Eq.~(\ref{delr}) are multiplied by 
\begin{equation}
\label{perpcor}
\left(1 - \frac{1}{2s} \left[1 - \frac 1N \sum_{\bf k} \frac{\epsilon_{\bf k}}{sJ_{\bf 0}}\right]\right) \approx 1 - \frac{0.08}{s}
\end{equation}
after taking into account first $1/s$-corrections. Meanwhile small interactions of the value of $\eta$ possibly could lead to a great renormalization of physical observable quantities at $|\omega|\ll\eta$. \cite{malold,syromyat} At the same time we derive ${\rm Im}\Delta_{\mu\nu}(\omega)$ at $|\omega|\gg\eta$. Thus we can use the expressions for $xx$- and $yy$- components obtained above within SWA.

We have to take into account also the longitudinal spin susceptibility $\langle s^z_{\bf k} s^z_{-\bf k}\rangle_\omega$. The first correction to it is of the first order of $1/s$. This quantity has been examined in Ref.~\cite{braune}. Nevertheless we derive now the corresponding expressions in a more convenient for us form. Using retarded Greens functions $g(\omega,{\bf k}) = \langle a_{\bf k}, a^\dagger_{\bf k} \rangle_\omega$, $f(\omega,{\bf k}) = \langle a_{\bf k}, a_{-\bf k} \rangle_\omega$, $\bar g(\omega,{\bf k}) = \langle a^\dagger_{-\bf k}, a_{-\bf k} \rangle_\omega = g^*(-\omega,-{\bf k})$ and $f^\dagger (\omega,{\bf k}) = \langle a^\dagger_{-\bf k}, a^\dagger_{\bf k} \rangle_\omega = f^*(-\omega,-{\bf k})$, where $a_{\bf k}$ and $a^\dagger_{\bf k}$ are operators of magnons creation and annihilation, we have
\begin{equation}
\label{gzz}
\langle s^z_{\bf k} s^z_{-\bf k}\rangle_\omega = -\frac T N\sum_{\omega_1}\sum_{{\bf k}_1 + {\bf k}_2 = {\bf k} + {\bf k}_0} 
[
f(i\omega_1,{\bf k}_1)f^\dagger(i\omega-i\omega_1,{\bf k}_2) + \bar g(i\omega_1,{\bf k}_1) g(i\omega-i\omega_1,{\bf k}_2)
].
\end{equation}
Functions $f$ and $g$ were calculated, e.g., in I. After simple evaluations one obtains
\begin{eqnarray}
\label{deltazz}
{\rm Im} \Delta_{zz}(\omega) = \frac{\pi}{2N^2} \sum_{{\bf k}_1, {\bf k}_2}&&
\frac{1-\cos([{\bf k}_1 + {\bf k}_2]{\bf R}_{12})}{\epsilon_{{\bf k}_1}\epsilon_{{\bf k}_2}}
\nonumber\\
&& {} \times \biggl\{ [1 + 2N(\epsilon_{{\bf k}_1})] \left[ (sJ_0)^2 + s^2J_{{\bf k}_1}J_{{\bf k}_2} - \epsilon_{{\bf k}_1}\epsilon_{{\bf k}_2} \right] [\delta(\omega + \epsilon_{{\bf k}_1} + \epsilon_{{\bf k}_2}) - \delta(\omega - \epsilon_{{\bf k}_1} - \epsilon_{{\bf k}_2})] \nonumber\\
&& {} +
2 \left[ N(\epsilon_{{\bf k}_2}) - N(\epsilon_{{\bf k}_1}) \right] \left[ (sJ_0)^2 + s^2J_{{\bf k}_1}J_{{\bf k}_2} + \epsilon_{{\bf k}_1}\epsilon_{{\bf k}_2} \right] \delta(\omega + \epsilon_{{\bf k}_1} - \epsilon_{{\bf k}_2})
\biggr\},
\end{eqnarray}
where the summations are over {\it chemical} Brillouin zone. The first term in the curly brackets in Eq.~(\ref{deltazz}) corresponds to emission or absorption of two magnons whereas the last term describes scattering of one spin wave. Let us discuss firstly the case of $T=0$. The Planck's functions in Eq.~(\ref{deltazz}) are zero and ${\rm Im} \Delta_{zz}(\omega)$ is proportional to $\omega^3$. Therefore, it is small in comparison with $xx$- and $yy$- components obtained above. 

At finite temperatures we have two important contributions to ${\rm Im} \Delta_{zz}(\omega)$. One of them arises at $|\omega|\ll T$ and originates from the second term in the curly brackets in Eq.~(\ref{deltazz}). It can be brought to the form $-\omega T^2 B/(s\Theta^3)$, where
\begin{equation}
\label{b}
B = \frac{24 \sqrt\pi}{J} \int_0^\infty dx \frac{x^2e^x}{(e^x - 1)^2} = \frac{8\pi^{5/2}}{J}.
\end{equation}
The second contribution originates from both terms in the curly brackets in Eq.~(\ref{deltazz}) and has the form $-A\omega^2 {\rm sgn}(\omega) v(T)/(s\Theta^2)$, where $v(T) = 8sJ/N \sum_{\bf k} N(\epsilon_{\bf k})/\epsilon_{\bf k}$. The logarithmic infra-red singularity in this expression is screened by the interaction stabilizing the long range order at $T\ne0$: $v(T) \cong (8/\sqrt\pi) (T/\Theta) \ln(T/[s\eta])$. There is no parameter screening this singularity in isotropic 2D AF. This singularity signifies that spin-wave approach does not suite for discussing the impurity dynamics in isotropic 2D AF.

It should be stressed that higher order $1/s$-corrections can give contributions proportional to products of $\omega^2$ ($\omega T^2$) and powers of $[T/(s\Theta)] \ln(T/[s\eta])$. As a result the obtained above correction proportional to $\omega T^2$ should be multiplied by a function $r(T)$. This function as well as $v(T)$ appear to be series in $[T/(s\Theta)] \ln(T/[s\eta])$. To find general expression for $r(T)$ and $v(T)$ is out of the scope of the present paper. Thus, we lead to $zz$-component of ${\rm Im} \Delta_{\mu\nu}(\omega)$ in Eq.~(\ref{delta}).

Notice that in the case of impurity coupled to one host spin, terms in Eq.~(\ref{delr}) containing $\cos({\bf k R}_{12})$ should be discarded and the spectral function appears to be proportional to a constant within SWA. There is no large corrections to this constant from higher order $1/s$-terms. Hence, one could not expect that longitudinal fluctuations play any remarkable role in this case. Then, we confirm that the so-called $T$-matrix approach \cite{iz,wan,chern} is appropriate for asymmetrically coupled defects. Recall that it bases on Dyson-Maleev transformation and deals with only bilinear part of the Hamiltonian.

\section{Calculation of the impurity susceptibility}
\label{gameval}

We present in this appendix some details of the impurity dynamical susceptibility calculation. We use for this general expression (\ref{chi}) and Eqs.~(\ref{gf}), (\ref{g++}) and (\ref{gam+-}) for the Green's function and the branches of the vertex. The expression for $\chi_P(\omega)$ is derived up to the order of $g^2$. 

According to Eqs.~(\ref{chi}), (\ref{g++}) and (\ref{gam+-}) the dynamical susceptibility can be represented as a sum of three components. The first one, $\chi_1(\omega)$, originates from Eq.~(\ref{chi}) as a result of replacement of the vertex by unity. The second, $\chi_2(\omega)$, appears from $f^2$-terms in Eqs.~(\ref{g++}) and (\ref{gam+-}). The third, $\chi_3(\omega)$, is a result of replacement of the vertex by the third term from Eq.~(\ref{gam+-}). Calculations of these three quantities are similar to those performed in I. Then we discuss the results only.

For $\chi_1(\omega)$ we obtain
\begin{eqnarray}
\label{chi1}
&&\chi_1(\omega) = \nonumber\\
&&
\frac{e^{-\lambda/T}}{\cal N}
\sum_{n,m}
|P_{nm}|^2
\left(
\frac{e^{(c_m+i\varGamma_{0m})/T} - e^{(c_n-i\varGamma_{0n})/T} }{\omega + c_m - c_n + i\varGamma_{0n} + i\varGamma_{0m}}
\left[
1 -
\left(R_{\Sigma n}^{(1)} + R_{\Sigma m}^{(1)}\right)\frac{f^2}{2\pi\Theta} \int_{-\infty}^\infty dx \frac{|x|[N(x) - N(-x)]\Lambda(x)}{x + \omega + c_m - c_n + i\varGamma_{0n} + i\varGamma_{0m}}
\right]
\right.\nonumber\\
&&{}+ \left.
\left(R_{\Sigma n}^{(1)} + R_{\Sigma m}^{(1)}\right) \left(e^{c_n/T} + e^{c_m/T}\right) \frac{f^2}{2\pi\Theta}\int_{-\infty}^\infty dx \frac{{\rm sgn}(x)\Lambda(x)}{x+\omega + c_m - c_n + i\varGamma_{0n} + i\varGamma_{0m}}
\right).
\end{eqnarray}

For $\chi_2(\omega)$ one has
\begin{eqnarray}
\label{chi21}
\chi_2(\omega) &=& 
\frac{e^{-\lambda/T}}{\cal N}
\sum_{n,m,q,l}
\frac{2{\rm Re}(P_{nm}S_{mq}^\nu P_{ql} S_{ln}^\mu d_{\mu\nu})}{\omega + c_m - c_n + i\varGamma_{0n} + i\varGamma_{0m}}\frac{f^2}{\pi\Theta}
\int_{-\infty}^\infty dx |x|xN(x)\Lambda(x) 
\nonumber\\
&&\times{}
\left[
\frac{e^{(c_m+i\varGamma_{om})/T}}{ (x-\omega+c_l-c_m-i\varGamma_{0l}-i\varGamma_{0m}) (x+c_q-c_m+i\varGamma_{0q}-i\varGamma_{0m})}
\right.\nonumber\\
&&{}-
\left.
\frac{e^{(c_n-i\varGamma_{on})/T}}{ (x+\omega+c_q-c_n+i\varGamma_{0q}+i\varGamma_{0n}) (x+c_l-c_n+i\varGamma_{0n}-i\varGamma_{0l})}
\right].
\end{eqnarray}
Eq.~(\ref{chi21}) has a simple form at $|\omega|\gg \tilde {\cal C}(T)g^2S/J$:
\begin{eqnarray}
\label{chi23}
\chi_2(\omega) &=& 
-\frac{e^{-\lambda/T}}{\cal N}
\sum_{n,m,q,l}
{\rm Re}(P_{nm}S_{mq}^\nu P_{ql} S_{ln}^\mu d_{\mu\nu})
\left(
e^{c_n/T} + e^{c_m/T}
\right)
\frac{f^2}{\pi\Theta}\int_{-\infty}^\infty dx \frac{{\rm sgn}(x)\Lambda(x)}{x + \omega + i\delta} .
\end{eqnarray}

For $\chi_3(\omega)$ one obtains
\begin{equation}
\label{chi3}
\chi_3(\omega) = 
\frac{e^{-\lambda/T}}{\cal N}
\sum_{n,m}
|P_{nm}|^2
\left(
\frac{e^{(c_m+2i\varGamma_{nm}-i\varGamma_{0n})/T} - e^{(c_n-i\varGamma_{0n})/T}}{\omega + c_m - c_n + 2i\varGamma_{nm}}
-
\frac{e^{(c_m+i\varGamma_{0m})/T} - e^{(c_n-i\varGamma_{0n})/T}}{\omega + c_m - c_n + i\varGamma_{0n} + i\varGamma_{0m}}
\right).
\end{equation}
It should be stressed that we use while calculating Eq.~(\ref{chi3}) that $\omega$ is close to one of the resonance frequencies $\omega_{res}^{(n)}$ given by Eq.~(\ref{ores}): $\left||\omega|-\omega_{res}^{(n)}\right|\ll T$. Meanwhile we can use expression (\ref{chi3}) at all $\omega$ as it is much smaller than $\chi_1(\omega)$ far from the resonances. 

It should be noted that the resonant terms in $\chi_{1,2}(\omega)$ and $\chi_3(\omega)$ are calculated in the order of $g^2$ and $g^0$, respectively. We would like to stress that calculation of $\chi_3(\omega)$ in higher orders demands taking into account in equation (\ref{g+-}) for $\Gamma^{+-}_{Pmn}(x+\omega,x)$ not only the most singular diagrams in each order of $g^2$. Their analysis is a cumbersome task that is out of the scope of the present paper. Thus we restrict ourself in this paper by calculation of resonant terms in the dynamical susceptibility in the order of $g^0$ and discard the resonant terms in $\chi_1(\omega)$ and $\chi_2(\omega)$ that are of the order of $g^2$. Notice that one can keep these terms while calculating static susceptibility because $\chi_3(0)=0$ and the problem of the cumbersome analysis of the higher order diagrams does not arise.
\footnote{One can makes sure that $\chi_3(0)=0$ using general expression (\ref{chi}) for $\chi_P(\omega)$. Expression (\ref{chi3}) for $\chi_3(\omega)$ is not equal to zero at $\omega=0$ that is a consequence of approximations made while evaluating Eq.~(\ref{chi3}). Meantime $\chi_3(0)$ given by Eq.~(\ref{chi3}) is much smaller than $\chi_1(0)$ and can be discarded in accordance with consideration presented below Eq.~(\ref{chi3}).}

%\bibliography{defects}

\begin{thebibliography}{35}
\expandafter\ifx\csname natexlab\endcsname\relax\def\natexlab#1{#1}\fi
\expandafter\ifx\csname bibnamefont\endcsname\relax
  \def\bibnamefont#1{#1}\fi
\expandafter\ifx\csname bibfnamefont\endcsname\relax
  \def\bibfnamefont#1{#1}\fi
\expandafter\ifx\csname citenamefont\endcsname\relax
  \def\citenamefont#1{#1}\fi
\expandafter\ifx\csname url\endcsname\relax
  \def\url#1{\texttt{#1}}\fi
\expandafter\ifx\csname urlprefix\endcsname\relax\def\urlprefix{URL }\fi
\providecommand{\bibinfo}[2]{#2}
\providecommand{\eprint}[2][]{\url{#2}}

\bibitem[{\citenamefont{Wan et~al.}(1993)\citenamefont{Wan, Harris, and
  Kumar}}]{wan}
\bibinfo{author}{\bibfnamefont{C.~C.} \bibnamefont{Wan}},
  \bibinfo{author}{\bibfnamefont{A.~B.} \bibnamefont{Harris}},
  \bibnamefont{and} \bibinfo{author}{\bibfnamefont{D.}~\bibnamefont{Kumar}},
  \bibinfo{journal}{Phys. Rev. B} \textbf{\bibinfo{volume}{48}},
  \bibinfo{pages}{1036} (\bibinfo{year}{1993}).

\bibitem[{\citenamefont{Hoglund and Sandvik}(2003)}]{hog1}
\bibinfo{author}{\bibfnamefont{K.~H.} \bibnamefont{Hoglund}} \bibnamefont{and}
  \bibinfo{author}{\bibfnamefont{A.~W.} \bibnamefont{Sandvik}},
  \bibinfo{journal}{Phys. Rev. Lett.} \textbf{\bibinfo{volume}{91}},
  \bibinfo{pages}{077204} (\bibinfo{year}{2003}).

\bibitem[{\citenamefont{Hoglund and Sandvik}(2004)}]{hog2}
\bibinfo{author}{\bibfnamefont{K.~H.} \bibnamefont{Hoglund}} \bibnamefont{and}
  \bibinfo{author}{\bibfnamefont{A.~W.} \bibnamefont{Sandvik}},
  \bibinfo{journal}{Phys. Rev. B} \textbf{\bibinfo{volume}{70}},
  \bibinfo{pages}{024406} (\bibinfo{year}{2004}).

\bibitem[{\citenamefont{Sushkov}(2000)}]{sush1}
\bibinfo{author}{\bibfnamefont{O.~P.} \bibnamefont{Sushkov}},
  \bibinfo{journal}{Phys. Rev. B} \textbf{\bibinfo{volume}{62}},
  \bibinfo{pages}{12135} (\bibinfo{year}{2000}).

\bibitem[{\citenamefont{Sushkov}(2003)}]{sush2}
\bibinfo{author}{\bibfnamefont{O.~P.} \bibnamefont{Sushkov}},
  \bibinfo{journal}{Phys. Rev. B} \textbf{\bibinfo{volume}{68}},
  \bibinfo{pages}{094426} (\bibinfo{year}{2003}).

\bibitem[{\citenamefont{Sachdev and Vojta}(2003)}]{sachdev}
\bibinfo{author}{\bibfnamefont{S.}~\bibnamefont{Sachdev}} \bibnamefont{and}
  \bibinfo{author}{\bibfnamefont{M.}~\bibnamefont{Vojta}},
  \bibinfo{journal}{Phys. Rev. B} \textbf{\bibinfo{volume}{68}},
  \bibinfo{pages}{064419} (\bibinfo{year}{2003}).

\bibitem[{\citenamefont{Vojta et~al.}(2000)\citenamefont{Vojta, Buragohain, and
  Sachdev}}]{vojta}
\bibinfo{author}{\bibfnamefont{M.}~\bibnamefont{Vojta}},
  \bibinfo{author}{\bibfnamefont{C.}~\bibnamefont{Buragohain}},
  \bibnamefont{and} \bibinfo{author}{\bibfnamefont{S.}~\bibnamefont{Sachdev}},
  \bibinfo{journal}{Phys. Rev. B} \textbf{\bibinfo{volume}{61}},
  \bibinfo{pages}{15152} (\bibinfo{year}{2000}).

\bibitem[{\citenamefont{Nagaosa et~al.}(1989)\citenamefont{Nagaosa, Hatsugai,
  and Imada}}]{nagaosa}
\bibinfo{author}{\bibfnamefont{N.}~\bibnamefont{Nagaosa}},
  \bibinfo{author}{\bibfnamefont{Y.}~\bibnamefont{Hatsugai}}, \bibnamefont{and}
  \bibinfo{author}{\bibfnamefont{M.}~\bibnamefont{Imada}}, \bibinfo{journal}{J.
  Phys. Soc. Jpn.} \textbf{\bibinfo{volume}{58}}, \bibinfo{pages}{978}
  (\bibinfo{year}{1989}).

\bibitem[{\citenamefont{Igarashi et~al.}(1995)\citenamefont{Igarashi, Murayama,
  and Fulde}}]{igar}
\bibinfo{author}{\bibfnamefont{J.}~\bibnamefont{Igarashi}},
  \bibinfo{author}{\bibfnamefont{K.}~\bibnamefont{Murayama}}, \bibnamefont{and}
  \bibinfo{author}{\bibfnamefont{P.}~\bibnamefont{Fulde}},
  \bibinfo{journal}{Phys. Rev. B} \textbf{\bibinfo{volume}{52}},
  \bibinfo{pages}{15966} (\bibinfo{year}{1995}).

\bibitem[{\citenamefont{Murayama and Igarashi}(1996)}]{murayama}
\bibinfo{author}{\bibfnamefont{K.}~\bibnamefont{Murayama}} \bibnamefont{and}
  \bibinfo{author}{\bibfnamefont{J.}~\bibnamefont{Igarashi}},
  \bibinfo{journal}{J. Phys. Soc. Jpn.} \textbf{\bibinfo{volume}{66}},
  \bibinfo{pages}{1157} (\bibinfo{year}{1996}).

\bibitem[{\citenamefont{Clarke et~al.}(1993)\citenamefont{Clarke, Giamarchi,
  and Shraiman}}]{clarke}
\bibinfo{author}{\bibfnamefont{D.~G.} \bibnamefont{Clarke}},
  \bibinfo{author}{\bibfnamefont{T.}~\bibnamefont{Giamarchi}},
  \bibnamefont{and} \bibinfo{author}{\bibfnamefont{B.~I.}
  \bibnamefont{Shraiman}}, \bibinfo{journal}{Phys. Rev. B}
  \textbf{\bibinfo{volume}{48}}, \bibinfo{pages}{7070} (\bibinfo{year}{1993}).

\bibitem[{\citenamefont{Oitmaa et~al.}(1995)\citenamefont{Oitmaa, Betts, and
  Aydin}}]{oitmaa}
\bibinfo{author}{\bibfnamefont{J.}~\bibnamefont{Oitmaa}},
  \bibinfo{author}{\bibfnamefont{D.~D.} \bibnamefont{Betts}}, \bibnamefont{and}
  \bibinfo{author}{\bibfnamefont{M.}~\bibnamefont{Aydin}},
  \bibinfo{journal}{Phys. Rev. B} \textbf{\bibinfo{volume}{51}},
  \bibinfo{pages}{2896} (\bibinfo{year}{1995}).

\bibitem[{\citenamefont{Kotov et~al.}(1998)\citenamefont{Kotov, Oitmaa, and
  Sushkov}}]{kot2}
\bibinfo{author}{\bibfnamefont{V.~N.} \bibnamefont{Kotov}},
  \bibinfo{author}{\bibfnamefont{J.}~\bibnamefont{Oitmaa}}, \bibnamefont{and}
  \bibinfo{author}{\bibfnamefont{O.}~\bibnamefont{Sushkov}},
  \bibinfo{journal}{Phys. Rev. B} \textbf{\bibinfo{volume}{58}},
  \bibinfo{pages}{8500} (\bibinfo{year}{1998}).

\bibitem[{\citenamefont{Syromyatnikov and Maleyev}(2005)}]{i}
\bibinfo{author}{\bibfnamefont{A.~V.} \bibnamefont{Syromyatnikov}}
  \bibnamefont{and} \bibinfo{author}{\bibfnamefont{S.~V.}
  \bibnamefont{Maleyev}}, \bibinfo{journal}{Phys. Rev. B}
  \textbf{\bibinfo{volume}{72}}, \bibinfo{pages}{174419}
  (\bibinfo{year}{2005}).

\bibitem[{\citenamefont{Abrikosov}(1965)}]{abrikos}
\bibinfo{author}{\bibfnamefont{A.~A.} \bibnamefont{Abrikosov}},
  \bibinfo{journal}{Physica} \textbf{\bibinfo{volume}{2}}, \bibinfo{pages}{5}
  (\bibinfo{year}{1965}).

\bibitem[{\citenamefont{Chernyshov et~al.}(2002)\citenamefont{Chernyshov, Chen,
  and Neto}}]{chern}
\bibinfo{author}{\bibfnamefont{A.~L.} \bibnamefont{Chernyshov}},
  \bibinfo{author}{\bibfnamefont{Y.~C.} \bibnamefont{Chen}}, \bibnamefont{and}
  \bibinfo{author}{\bibfnamefont{A.~H.~C.} \bibnamefont{Neto}},
  \bibinfo{journal}{Phys. Rev. B} \textbf{\bibinfo{volume}{65}},
  \bibinfo{pages}{104407} (\bibinfo{year}{2002}).

\bibitem[{\citenamefont{Izyumov and Medvedev}(1973)}]{iz}
\bibinfo{author}{\bibfnamefont{Y.~A.} \bibnamefont{Izyumov}} \bibnamefont{and}
  \bibinfo{author}{\bibfnamefont{M.~V.} \bibnamefont{Medvedev}},
  \emph{\bibinfo{title}{Magnetically Ordered Crystals Containing Impurities}}
  (\bibinfo{publisher}{Consultants Bureau}, \bibinfo{address}{New York},
  \bibinfo{year}{1973}).

\bibitem[{\citenamefont{Shender}(1982)}]{shender}
\bibinfo{author}{\bibfnamefont{E.~F.} \bibnamefont{Shender}},
  \bibinfo{journal}{Sov. Phys. JETP} \textbf{\bibinfo{volume}{56}},
  \bibinfo{pages}{178} (\bibinfo{year}{1982}).

\bibitem[{\citenamefont{Ivanov}(1972)}]{ivanov}
\bibinfo{author}{\bibfnamefont{M.~A.} \bibnamefont{Ivanov}},
  \bibinfo{journal}{Fiz. Tverd. Tela} \textbf{\bibinfo{volume}{14}},
  \bibinfo{pages}{562} (\bibinfo{year}{1972}).

\bibitem[{\citenamefont{Maleyev}(1979)}]{mal2}
\bibinfo{author}{\bibfnamefont{S.~V.} \bibnamefont{Maleyev}},
  \bibinfo{journal}{Sov. Phys. JETP} \textbf{\bibinfo{volume}{52}},
  \bibinfo{pages}{1008} (\bibinfo{year}{1979}).

\bibitem[{\citenamefont{Maleyev}(1983)}]{mal4}
\bibinfo{author}{\bibfnamefont{S.~V.} \bibnamefont{Maleyev}},
  \bibinfo{journal}{Sov. Phys. JETP} \textbf{\bibinfo{volume}{57}},
  \bibinfo{pages}{149} (\bibinfo{year}{1983}).

\bibitem[{\citenamefont{Maleyev}(1994)}]{mal5}
\bibinfo{author}{\bibfnamefont{S.~V.} \bibnamefont{Maleyev}},
  \bibinfo{journal}{Phys. Rev. B} \textbf{\bibinfo{volume}{50}},
  \bibinfo{pages}{302} (\bibinfo{year}{1994}).

\bibitem[{\citenamefont{Lovesey}(1968{\natexlab{a}})}]{lov1}
\bibinfo{author}{\bibfnamefont{S.~W.} \bibnamefont{Lovesey}},
  \bibinfo{journal}{J. Phys. C (Proc. Phys. Soc.)}
  \textbf{\bibinfo{volume}{1}}, \bibinfo{pages}{102}
  (\bibinfo{year}{1968}{\natexlab{a}}).

\bibitem[{\citenamefont{Lovesey}(1968{\natexlab{b}})}]{lov2}
\bibinfo{author}{\bibfnamefont{S.~W.} \bibnamefont{Lovesey}},
  \bibinfo{journal}{J. Phys. C (Proc. Phys. Soc.)}
  \textbf{\bibinfo{volume}{1}}, \bibinfo{pages}{118}
  (\bibinfo{year}{1968}{\natexlab{b}}).

\bibitem[{\citenamefont{Kopietz}(1990)}]{kop}
\bibinfo{author}{\bibfnamefont{P.}~\bibnamefont{Kopietz}},
  \bibinfo{journal}{Phys. Rev. B} \textbf{\bibinfo{volume}{41}},
  \bibinfo{pages}{9228} (\bibinfo{year}{1990}).

\bibitem[{\citenamefont{Chakravarty et~al.}(1989)\citenamefont{Chakravarty,
  Halperin, and Nelson}}]{chak}
\bibinfo{author}{\bibfnamefont{S.}~\bibnamefont{Chakravarty}},
  \bibinfo{author}{\bibfnamefont{B.~I.} \bibnamefont{Halperin}},
  \bibnamefont{and} \bibinfo{author}{\bibfnamefont{D.~R.}
  \bibnamefont{Nelson}}, \bibinfo{journal}{Phys. Rev. B}
  \textbf{\bibinfo{volume}{39}}, \bibinfo{pages}{2344} (\bibinfo{year}{1989}).

\bibitem[{\citenamefont{Ty\v{c} and Halperin}(1990)}]{tyc}
\bibinfo{author}{\bibfnamefont{S.}~\bibnamefont{Ty\v{c}}} \bibnamefont{and}
  \bibinfo{author}{\bibfnamefont{B.~I.} \bibnamefont{Halperin}},
  \bibinfo{journal}{Phys. Rev. B} \textbf{\bibinfo{volume}{42}},
  \bibinfo{pages}{2096} (\bibinfo{year}{1990}).

\bibitem[{\citenamefont{Thurber et~al.}(1997)\citenamefont{Thurber, Hunt, Imai,
  Chou, and Lee}}]{thur}
\bibinfo{author}{\bibfnamefont{K.~R.} \bibnamefont{Thurber}},
  \bibinfo{author}{\bibfnamefont{A.~W.} \bibnamefont{Hunt}},
  \bibinfo{author}{\bibfnamefont{T.}~\bibnamefont{Imai}},
  \bibinfo{author}{\bibfnamefont{F.~C.} \bibnamefont{Chou}}, \bibnamefont{and}
  \bibinfo{author}{\bibfnamefont{Y.~S.} \bibnamefont{Lee}},
  \bibinfo{journal}{Phys. Rev. Lett.} \textbf{\bibinfo{volume}{79}},
  \bibinfo{pages}{171} (\bibinfo{year}{1997}).

\bibitem[{\citenamefont{Castilla and Chakravarty}(1991)}]{cast}
\bibinfo{author}{\bibfnamefont{G.~E.} \bibnamefont{Castilla}} \bibnamefont{and}
  \bibinfo{author}{\bibfnamefont{S.}~\bibnamefont{Chakravarty}},
  \bibinfo{journal}{Phys. Rev. B} \textbf{\bibinfo{volume}{43}},
  \bibinfo{pages}{13687} (\bibinfo{year}{1991}).

\bibitem[{\citenamefont{Canali et~al.}(1992)\citenamefont{Canali, Girvin, and
  Wallin}}]{canali}
\bibinfo{author}{\bibfnamefont{C.~M.} \bibnamefont{Canali}},
  \bibinfo{author}{\bibfnamefont{S.~M.} \bibnamefont{Girvin}},
  \bibnamefont{and} \bibinfo{author}{\bibfnamefont{M.}~\bibnamefont{Wallin}},
  \bibinfo{journal}{Phys. Rev. B} \textbf{\bibinfo{volume}{45}},
  \bibinfo{pages}{10131} (\bibinfo{year}{1992}).

\bibitem[{\citenamefont{Igarashi}(1992)}]{igar2}
\bibinfo{author}{\bibfnamefont{J.-I.} \bibnamefont{Igarashi}},
  \bibinfo{journal}{Phys. Rev. B} \textbf{\bibinfo{volume}{46}},
  \bibinfo{pages}{10763} (\bibinfo{year}{1992}).

\bibitem[{\citenamefont{Harris et~al.}(1971)\citenamefont{Harris, Kumar,
  Halperin, and Hohenberg}}]{harris}
\bibinfo{author}{\bibfnamefont{A.~B.} \bibnamefont{Harris}},
  \bibinfo{author}{\bibfnamefont{D.}~\bibnamefont{Kumar}},
  \bibinfo{author}{\bibfnamefont{B.~I.} \bibnamefont{Halperin}},
  \bibnamefont{and} \bibinfo{author}{\bibfnamefont{P.~C.}
  \bibnamefont{Hohenberg}}, \bibinfo{journal}{Phys. Rev. B}
  \textbf{\bibinfo{volume}{3}}, \bibinfo{pages}{961} (\bibinfo{year}{1971}).

\bibitem[{\citenamefont{Maleyev}(2000)}]{malold}
\bibinfo{author}{\bibfnamefont{S.}~\bibnamefont{Maleyev}},
  \bibinfo{journal}{Phys. Rev. Lett.} \textbf{\bibinfo{volume}{85}},
  \bibinfo{pages}{3281} (\bibinfo{year}{2000}).

\bibitem[{\citenamefont{Syromyatnikov and Maleyev}(2001)}]{syromyat}
\bibinfo{author}{\bibfnamefont{A.~V.} \bibnamefont{Syromyatnikov}}
  \bibnamefont{and} \bibinfo{author}{\bibfnamefont{S.~V.}
  \bibnamefont{Maleyev}}, \bibinfo{journal}{Phys. Rev. B}
  \textbf{\bibinfo{volume}{65}}, \bibinfo{pages}{012401}
  (\bibinfo{year}{2001}).

\bibitem[{\citenamefont{Braune and Maleyev}(1990)}]{braune}
\bibinfo{author}{\bibfnamefont{S.}~\bibnamefont{Braune}} \bibnamefont{and}
  \bibinfo{author}{\bibfnamefont{S.~V.} \bibnamefont{Maleyev}},
  \bibinfo{journal}{Z. Phys. B} \textbf{\bibinfo{volume}{81}},
  \bibinfo{pages}{69} (\bibinfo{year}{1990}).

\end{thebibliography}

\newpage

\begin{table}
\caption{Renormalization of spin-wave damping and velocity by interaction with spin-$\frac12$ impurities in two regimes: $|\omega|\gg\omega_0$ and $|\omega|\ll\omega_0$, where $\omega_0$ is given by Eq.~(\ref{omlim}). Ranges of validity of the theory in each regime are also presented (see the text).
\label{specren}
}
\begin{ruledtabular}
\begin{tabular}{|c|c|c|}
	& $|\omega|\gg\omega_0$ & $|\omega|\ll\omega_0$ \\
		\hline
spin-wave damping & $|\omega|\frac{\displaystyle n f^4}{\displaystyle 2\pi} R_\chi u({\bf k})$ & $\frac{\displaystyle \Theta}{\displaystyle T} \frac{\displaystyle \omega^2 \varGamma}{\displaystyle \omega^2 + 4\varGamma^2} \frac{\displaystyle n f^2}{\displaystyle 8\pi}u({\bf k})$ \\
spin-wave velocity & $c \sqrt{ 1 - \ln\left|\frac{\displaystyle \Theta}{\displaystyle \omega}\right| \frac{\displaystyle 2 n f^4}{\displaystyle \pi^2} R_\chi u({\bf k}) }$ & $c \sqrt{ 1 - \frac{\displaystyle \Theta}{\displaystyle T} \frac{\displaystyle \varGamma^2}{\displaystyle \omega^2 + 4\varGamma^2} \frac{\displaystyle n f^2}{\displaystyle 2\pi}u({\bf k}) }$ \\
range of validity of the theory & $\frac{\displaystyle |\omega|}{\displaystyle \Theta} \gg 0.1 \sqrt{R_\chi} \sqrt n f^2$ & $\frac{\displaystyle |\omega|(\omega^2+4\varGamma^2)}{\displaystyle \Theta^3} \gg 0.004nf^2\frac{\displaystyle \varGamma}{\displaystyle T}$
\end{tabular}
\end{ruledtabular}
\end{table}

\end{document}